\documentclass{aa}  
\usepackage{natbib}
\usepackage{graphicx}
\usepackage{txfonts}
\usepackage{amsmath}
\usepackage{gensymb}
\usepackage{xcolor}
\newcommand{\edit}[1]{\textcolor{black}{#1}}

\newcommand{\editguy}[1]{\small \textcolor{black}{#1}}

\usepackage{amsmath,amsfonts,amssymb}

\newcommand{\V}[1]{\boldsymbol{#1}}
\newcommand{\Tag}[1]{{\mathsf{#1}}}
\newcommand{\Reals}{\mathbb{R}}
\newcommand{\Cost}{\mathcal{J}}

\begin{document}

   \title{Evidence for localized onset of episodic mass loss in \object{Mira}}


   \author{G. Perrin
          \inst{1}
          \and
          S.T. Ridgway\inst{1,2}
          \and
          S. Lacour \inst{1}
          \and
          X. Haubois\inst{3}
          \and
          \'E. Thi\'ebaut\inst{4}
          \and
          J. P. Berger\inst{5}
          \and
          M.G. Lacasse\inst{6}
          \and
          R. Millan-Gabet\inst{7}
          \and
          J. D. Monnier\inst{8}
          \and
          E. Pedretti\inst{9}
          \and
          S. Ragland\inst{10}
          \and 
          W. Traub\inst{11}
          }
          
 \institute{LESIA, Observatoire de Paris, Universit\'{e} PSL, CNRS, Sorbonne Universit\'{e}, Universit\'{e} de Paris, 5 place Jules Janssen, 92195 Meudon, France \and NSF's NOIRLab, 950 N. Cherry Ave., Tucson, AZ 85719, USA \and European Southern Observatory, Alonso de C\'ordova 3107, Vitacura, Casilla 19001, Santiago de Chile, Chile \and Universit\'{e} de Lyon, Universit\'e Lyon1, ENS de Lyon, CNRS, Centre de Recherche Astrophysique de Lyon UMR 5574, 69230 Saint-Genis-Laval, France \and Univ. Grenoble Alpes, CNRS, IPAG, F-38000 Grenoble, France \and Harvard-Smithsonian Center for Astrophysics, 60 Garden Street, Cambridge, MA 02138, USA \and  Infrared Processing and Analysis Center, California Institute of Technology, Pasadena, CA, 91125, USA \and University of Michigan, 941 Dennison Building, 500 Church Street, Ann Arbor, MI 48109-1090, USA \and RAL Space, STFC Rutherford Appleton Laboratory, Harwell, Didcot, OX11 0QX, UK \and W. M. Keck Observatory, 65-1120 Mamalahoa Hwy, Kamuela, HI 96743, USA \and Jet Propulsion Laboratory, California Institute of Technology, M/S 301-451, 4800 Oak Grove Dr., Pasadena CA, 91109, USA
                 }

   \date{Received 3 January 2020 / Accepted 17 August 2020 }

 
  \abstract
   {Mass loss from long-period variable stars (LPV) is an important contributor to the evolution of galactic abundances. Dust formation is understood to play an essential role in mass loss. It has, however, proven difficult to develop measurements that strongly constrain the location and timing of dust nucleation and acceleration.}
   {Interferometric imaging has the potential to constrain the geometry and dynamics of mass loss. High angular resolution studies of various types have shown that LPVs have a distinct core-halo structure. These have also shown that LPV images commonly exhibit a non-circular shape. The nature of this shape and its implications are yet to be understood.}
   {Multi-telescope interferometric measurements taken with the Interferometric Optical
Telescope Array (IOTA) provide imagery of the LPV Mira in the H-band. This wavelength region is well suited to studying mass loss given the low continuum opacity, which allows for emission to be observed over a very long path in the stellar atmosphere and envelope.}
   {The observed visibilities are consistent with a \editguy{simple} core-halo model \editguy{to represent the central object and the extended molecular layers} but, in addition, they demonstrate a substantial asymmetry. An analysis with \editguy{ image reconstruction software} shows that the asymmetry is consistent with a localized absorbing patch. The observed opacity is tentatively associated with small dust grains, which will grow substantially during a multi-year ejection process. Spatial information along with a deduced dust content of the cloud, known mass loss rates, and ejection velocities provide   evidence for the pulsational pumping of the \editguy{extended molecular layers}. The cloud may be understood as a spatially local zone of enhanced dust formation, very near to the pulsating halo. The observed mass loss could be provided by several such active regions around the star.}
  {This result provides an additional clue for better understanding the clumpiness of dust production in the atmosphere of AGB stars. It is compatible with scenarios where the combination of pulsation and convection play a key role in the process of mass loss.}

 \keywords {stars: atmospheres -  stars: AGB - stars: mass-loss - techniques: interferometric -  infrared: stars - stars: individual: Mira}
   \maketitle
%
\section{Introduction}
All stars of low to intermediate mass evolve through the AGB phase, with a dredging up of synthesized elements, and undergo extensive mass loss as long-period variables (LPVs).  This return of heavy elements to the interstellar medium constitutes a critical step in the evolution of stars and galaxies.  The LPV is a very complex phenomenon and our understanding of it remains incomplete.  Both modeling and observations continue to guide our progress in this area.

LPVs vary in brightness cyclically, with periods on the order of a few hundred days.  Photometric monitoring shows that these periodic variations are often slightly or even severely irregular.   The  improving capabilities at present for direct imagery of LPVs show that the LPV atmosphere is asymmetric, at least in some cases and potentially for all.  Speckle and other imaging techniques have been used to obtain high angular resolution information about LPVs with ground-based telescopes and interferometers, as well as with HST.  \citet{Scholz2003} gives an extensive list of such observations prior to 2003.  \citet{Creech-Eakman&Thompson2009} summarize a seven-year study of 100 LPVs using the Palomar Testbed Interferometer and report extensive results for apparent sizes and their variation with position angle and time. \citet{Ragland2006}, using the Interferometric Optical Telescope Array (IOTA), found that 1/3 of 56 nearby LPV stars surveyed showed asymmetry. The authors speculated that all of them would appear asymmetric if studied with sufficient resolution. This trend was confirmed by the observations of \citet{Wittkowski2011} at VLTI with AMBER which measured deviations from centro-symmetry on four Mira variables. Later, \citet{Wittkowski2016} used AMBER to observe large-scale inhomogeneities that contribute a few percent of the total flux in six Mira variables. The detailed structure of the inhomogeneities show a variability on time scales of three months and above.   This is compatible with recent 3D dynamical simulations of AGB stars which deal self-consistently with convection and pulsations \citep{Freytag2017}. Convection, together with pulsational shocks, contributes to the raising of material up to altitudes where dust can condense. This knowledge leads us to consider whether the spatial inhomogeneities revealed by observations and predicted by models could also lead to inhomogeneous zones of dust formation.

\citet{Ohnaka2016} used visible polarimetric imaging observations of W Hya taken with VLT/SPHERE-ZIMPOL to detect three clumpy dust clouds at two stellar radii. The authors deduced large dust grains (0.4-0.5 $\mu$m) of Al$_2$O$_3$, Mg$_2$SiO$_4$, or MgSiO$_3$. 
The grain size is consistent with the prediction of hydrodynamical models for mass loss driven by scattering due to Fe-free micron-sized grains \citep{Hofner2008}.
The detection of clumpy dust clouds close to the star lends support to dust formation induced by pulsation and large convective cells. Observations at similar spatial scales but at longer wavelengths in the millimeter range have also revealed asymmetries. \citet{Wong2016}  detected bright SiO-emitting clumps with ALMA in Mira at distances around two stellar radii. Also with ALMA on Mira, \citet{Kaminski2017} revealed the anisotropy of the distribution of TiO$_2$ in the atmosphere of Mira up to 5.5 stellar radii. This may suggest inhomogeneous distributions for both dust and its precursors near the photosphere, which is compatible with clumpy nucleation sites.

The shapes suggest a structure that may eventually contribute to the form of the planetary nebula (PN) expected to follow the AGB phase. However, PNe almost certainly represent the shape of the outer envelope and beyond, arising perhaps at some distance from the asymmetrical physical phenomena occurring at yet deeper layers which give rise to them. \editguy{In addition, part of the shape may be due to binarity.} Measurements with higher angular resolution at wavelengths that can penetrate deeply into the envelope promise to stand as the best opportunity to tie together the most reliable part of the LPV theory - the unobservable interior processes - with the complexity of the observed exterior.

Here, we report  the continuation of a series of LPV studies carried out at IOTA (\citet{Traub1998}, which   offers excellent (u,v) coverage, which can be enhanced with multi-spectral measurements.  In \citet{Perrin2004}, we showed strong evidence in favor of the presence of a very deep, but surprisingly transparent, extended  atmosphere that is analogous to, and perhaps similar to, the MOLsphere, which Tsuji ascribed to luminous, non-Mira stars. This \edit{interpretation was supported} by other observations, for example \cite{Lacour2009} and \citet{Le_Bouquin2009}. \edit{Since dynamical models reproduce the characteristics of Tsuji's MOLsphere, without, however, an actual discrete layer (\citet{Bladh2015}), the historical term is no longer understood as descriptively correct. We follow our referee's recommendation to describe this region as "extended molecular layers."}

In our previous studies, we utilized pair-wise combination of telescopes and only reported the analysis based on fringe visibilities and spherical models.
In this study, we use the IONIC beam combiner instrument of IOTA \citep{Berger2003}, which combines three telescopes simultaneously and measures closure phases. Squared visibilities and closure phases can be combined to reconstruct images. We present the observations and data reduction in Sect.~\ref{Sec:obs}. The data are first fit with a circularly symmetric model in Sect.~\ref{Sec:prior}. These data have allowed us to synthesize an image of the LPV Mira (omicron Ceti), presented in Sect.~\ref{Sec:imaging}, which is sufficiently indicative of the asymmetric structure to motivate inclusion of an asymmetric component in our models and to derive some related parameters in Sect.~\ref{Sec:model}.  We follow with a discussion of how this model relates to other measurements and how it may fit in the context of expected phenomena in LPV oscillation and mass loss in Sect.~\ref{Sec:discussion}.  Finally, we comment on the future of LPV imagery in Sect.~\ref{Sec:future}.


\section{Observations and data reduction}
\label{Sec:obs}
\begin{table*}
\caption{Log of the observations of Mira with IONIC on IOTA. The visual phase was computed from data from the AFOEV. The configurations give the positions of the three telescopes in meter with respect to the corner of the array.}

\centering                          
\begin{tabular}{l c l l}        
\hline\hline                 
Date (UT) & Visual phase & Configuration & Baseline lengths (m) \\    
\hline                        
   2005 October 5, 6, 7         & 0.41  & A5-B5-C0              & 5.00, 5.00, 7.07 \\
   2005 October 8               & 0.42  & A5-B15-C0             & 5.00, 15.00, 15.81   \\
   2005 October 10, 11  & 0.43  & A15-B15-C0            & 15.00, 15.00, 21.21 \\
   2005 October 12, 13  & 0.44  & A25-B15-C0            & 15.00, 25.00, 29.15 \\
   2005 October 15              & 0.45  & A30-B15-C15   & 15.00, 21.21, 33.54 \\
\hline                                   
\end{tabular}
\label{table:log}      
\end{table*}

\begin{table}
\caption{Reference stars used to calibrated the Mira data. Diameters are from \cite{Borde2002}. }
\label{table:calibrators}      
\centering                          
\begin{tabular}{l l l}        
\hline\hline                 
Star    & Spectral type & Uniform disk diameter (mas) \\    
\hline                        
HD 8512         & K0 IIIb       &       $2.69\pm0.030\,$mas \\
HD 16212                & M0 III        &       $3.02\pm0.032\,$mas \\
\hline                                   
\end{tabular}
\end{table}

\begin{figure}
   \centering
   \includegraphics[width=\hsize]{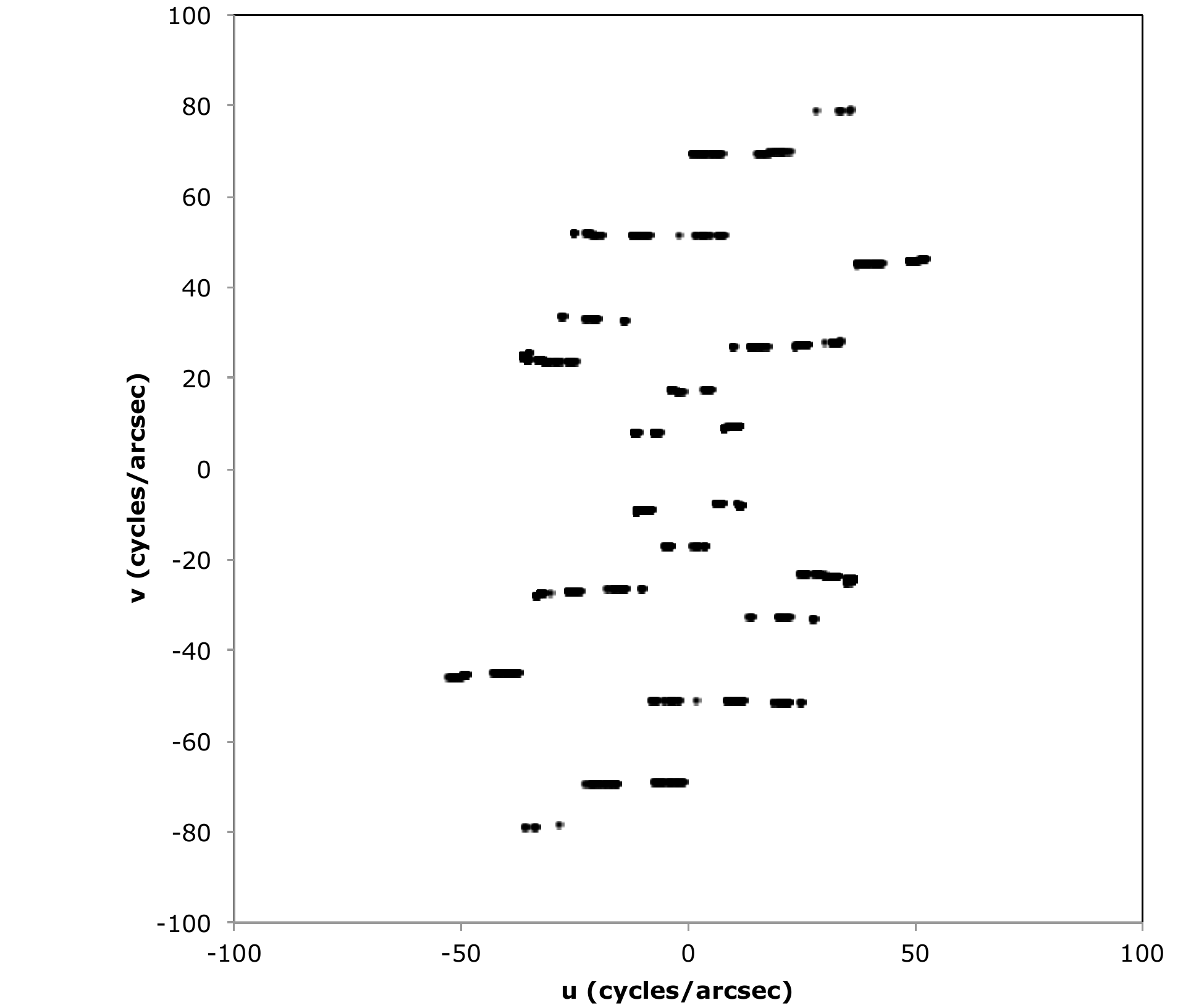}
      \caption{(u,v) coverage achieved on Mira during the observations with IOTA/IONIC. Each point marks an exposure.}
         \label{Fig:(u,v)}
   \end{figure}

 \begin{figure}
   \centering
   \includegraphics[width=\hsize]{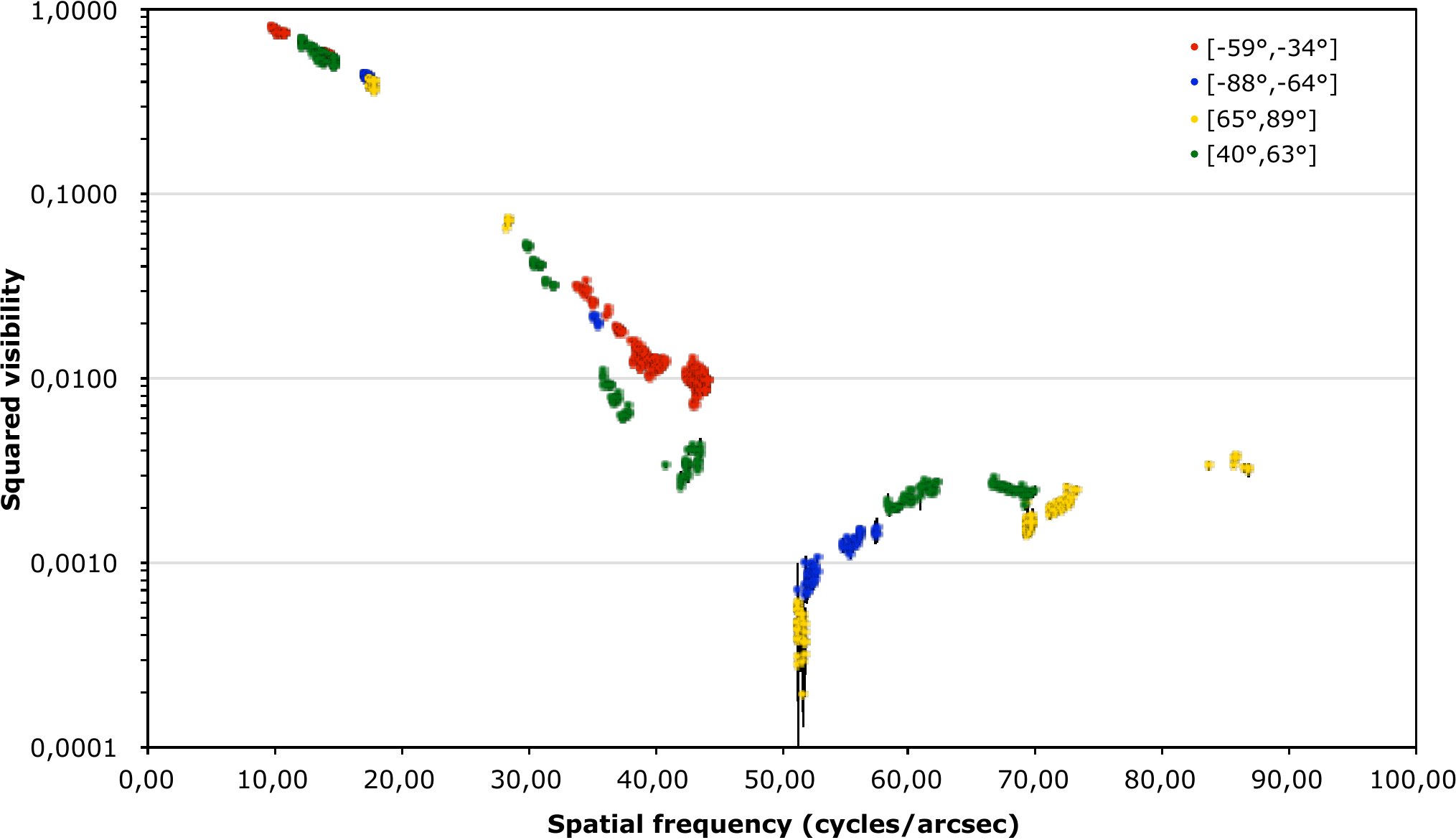}
      \caption{Plot of the squared visibilities of Mira measured with IOTA/IONIC vs. spatial frequency. The color code gives the orientation of the baselines arranged in four bins: [-88$\degree$,-64$\degree$] (blue), [-59$\degree$,-34$\degree$] (red), [40$\degree$,63$\degree$] (green), [65$\degree$,59$\degree$] (yellow). The color code directly shows that the source is not circularly symmetric.}
         \label{Fig:V2}
   \end{figure}

\begin{figure}
   \centering
   \includegraphics[width=\hsize]{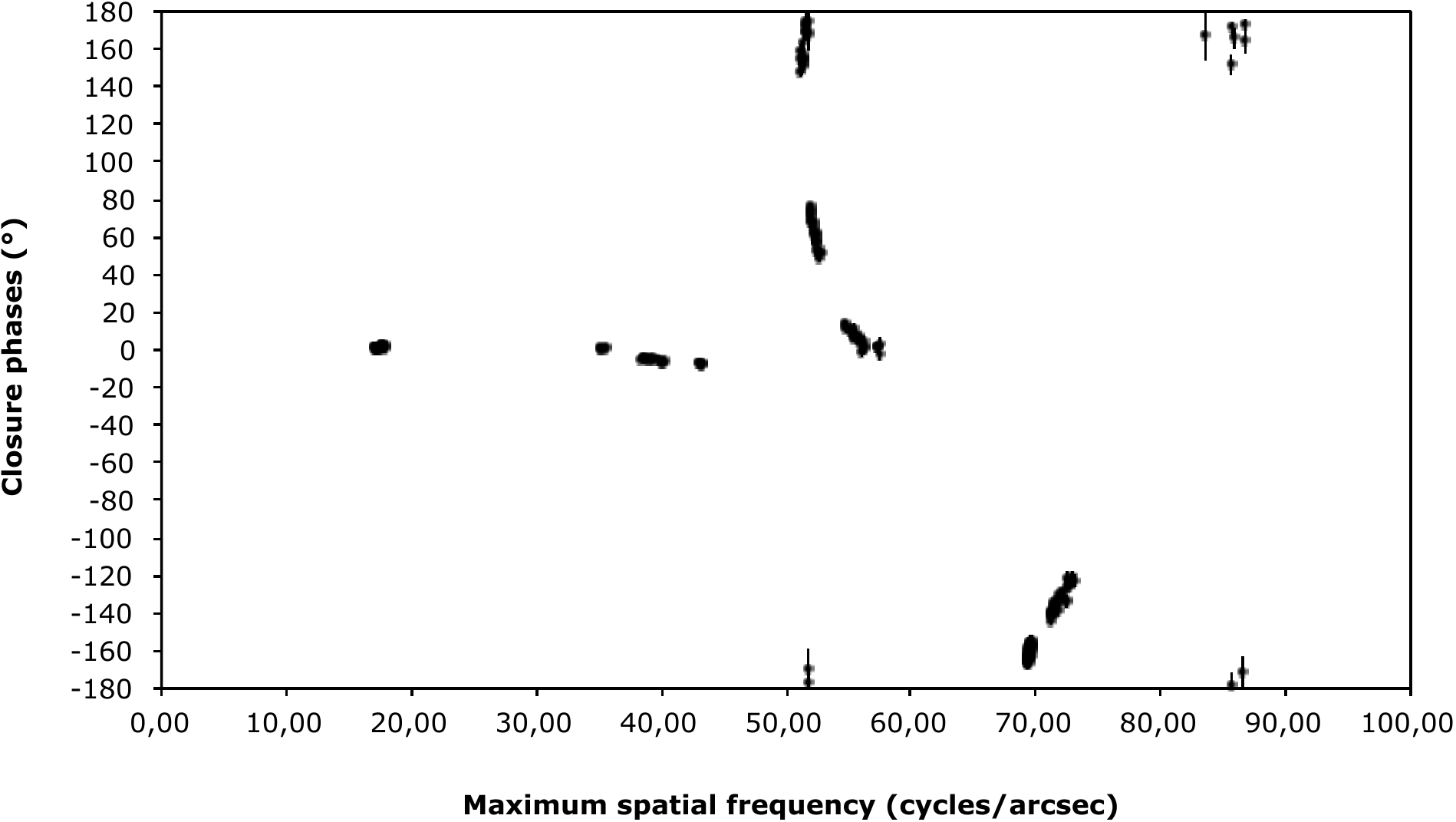}
      \caption{Plot of the closure phases of Mira measured with IOTA/IONIC. The closure phases are plotted against the maximum spatial frequency of each triangle. The closure phases are compatible with 0($\pi$) up to the first zero crossing of the squared visibilities near 50 cycles/arcsec. Beyond that spatial frequency, closure phases depart from 0($\pi$) meaning the source is not centro-symmetric at scales smaller than the typical radius of the star.}
         \label{Fig:CP}
   \end{figure}

\subsection{Description of IOTA observations}
The interferometric data presented herein were obtained using the IOTA  interferometer (Traub et al. 2003), which was a long baseline interferometer operating at near-infrared wavelengths until 2006. It consisted of three 0.45 m telescopes movable among 17 stations along two orthogonal linear arms. IOTA could synthesize a total aperture size of 35$\times$15 m, corresponding to an angular resolution of 10$\times$23 milliarcseconds at $1.65\,\mu m$. Visibility and closure phase measurements were obtained using the integrated optics combiner IONIC (Berger et al. 2003); light from the three telescopes is focused into single-mode fibers and injected into the planar integrated optics (IO) device. Six IO couplers allow recombinations between each pair of telescopes. Fringe detection was done using a Rockwell PICNIC detector (Pedretti et al. 2004). The interference fringes were temporally-modulated on the detector by scanning piezo mirrors placed in two of the three beams of the interferometer.


Observations were carried out in the H band ($1.5\,\mu$m $\leq \lambda \leq 1.8 \, \mu$m) divided into seven spectral channels. The science target observations were interleaved with identical observations of unresolved or partially resolved stars, used to calibrate the interferometer’s instrumental response and effects of atmospheric seeing on the visibility amplitudes. The calibrator sources were chosen from the \citet{Borde2002} catalog. Mira was observed in October 2005 over nine nights and using five different configurations of the interferometer. Full observation information can be found in Table~\ref{table:log}, including dates of observation, interferometer configurations, and visual phase computed from the Association Fran\c{c}aise des Observateurs d'\'Etoiles Variables (http://cdsarc.u-strasbg.fr/afoev/). The calibrators are listed in Table~\ref{table:calibrators}. Figure~\ref{Fig:(u,v)} shows the (u,v) coverage achieved during this observation run. The geometry of the IOTA interferometer and the position of the star on the sky constrained the frequency coverage, which is elongated and primarily located in the north-south direction. 

\subsection{Data reduction}
Reduction of the IONIC visibility data was carried out using custom software by \citet{Monnier2006}, which is similar in its main principles to those described by \citet{Coude1997}. The premise is to measure the power spectrum of each interferogram (proportional to the target squared visibility, $V^2$) after correcting for intensity fluctuations and subtracting bias terms due to read noise, residual intensity fluctuations, and photon noise \citep{Perrin2003}. Next a correction is applied by the data pipeline for the variable flux ratios for each baseline by using a flux transfer matrix \citep{Monnier2001}. Finally, raw squared visibilities are calibrated using the visibilities obtained by the same means on the calibrator stars. As in \citet{Lacour2008} less than a 1\% uncertainty in visibility is associated with the calibrators. This corresponds to a 2\% calibration error for $V^2$. Therefore, we systematically added a 2\% calibration error to all the calibrated squared visibilities in this paper. In order to measure the closure phase (CP), a real-time fringe tracking algorithm \citep{Pedretti2005} ensured that interference occurred simultaneously for all baselines. We required that interferograms are detected for at least two of the three baselines in order to assure a good closure phase measurement. This technique, called baseline bootstrapping, allowed for precise visibility and closure phase measurements for a third baseline with very small coherence fringe amplitude. We followed the method of \citet{Baldwin1996} to calculate the complex triple amplitudes and derive the closure phases. Pair-wise combiners (such as IONIC) can have a large instrumental offset for the closure phase which can be calibrated by the closure phase of the calibrator stars.

The squared visibilities and the closure phases are presented in Fig.~\ref{Fig:V2} and Fig.~\ref{Fig:CP}, respectively. The colors show different baseline orientations. Obviously the source is not circularly symmetric as the visibilities depend on azimuth. A zero crossing is visible in the visibility function near 50 cycles/arcsec, the signature of sharp edges for the main contributor to the flux. In addition, the closure phases depart from 0 modulo $\pi$ beyond the first zero of the visibility function, showing a departure from centro-symmetry at scales that are smaller or comparable to that of the central object. 


\section{Model fitting with a circularly symmetric model}
\label{Sec:prior}

The $V^{2}$ data were first compared to a model star photosphere + partially transparent shell \citep{Perrin2004} to derive the main parameters of the model: photosphere diameter, shell diameter, shell optical depth, shell temperature, and photosphere temperature. The degeneracy of the temperature fit is solved by fitting the flux. For this purpose, we used the phases from the AFOEV data, as shown in Table~\ref{table:log}. For the average photometric phase of 0.43 the magnitudes interpolated from \citet{Whitelock2000} are H (-1.7 to -1.8) and K (-2.2 to -2.3). We note that the infrared fluxes are much less variable than the visible fluxes. The H and K bands gave weak constraints on the photospheric temperature, probably because the photospheric diameter is not uniquely determined.  The flux gave a minimum value of 3000\,K, and it was set to 3400\,K, which is a rough average of the temperatures measured by  \citet{Perrin2004}  at earlier phases. This yields the global parameters for the model in Table~\ref{tab:prior}.

\begin{table}
\caption{Parameters of the star+shell model fit to the $V^2$ data.}

\begin{tabular}{l l l l l}        
\hline\hline                 
D$_{\star}$     & T$_{\star}$   & D$_{shell}$   & T$_{shell}$ & $\tau$ \\    
\hline                        
 21.1 mas               & 3400 K                & 40.00 mas     & 2000 K         & 0.70 \\
 \hline                                   
\end{tabular}
\label{tab:prior}
\end{table}


\section{Imaging of Mira}
\label{Sec:imaging}
\subsection{Image reconstruction}
\begin{figure}
   \centering
   \includegraphics[width=\columnwidth]{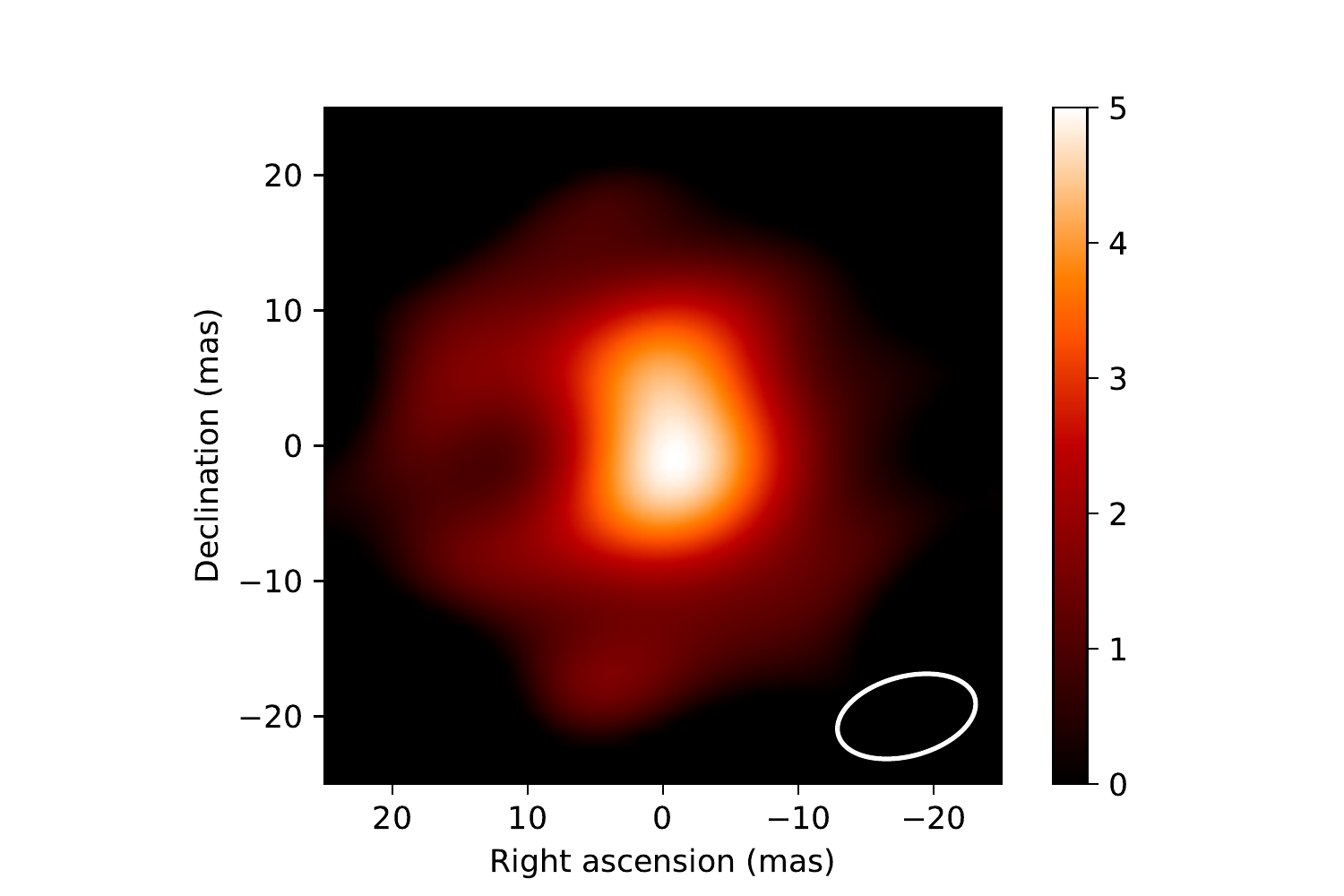}

   \caption{Image of Mira reconstructed with the MiRA software. \editguy{A dark patch is visible west of the bright central peak. The scale is linear with 0 intensity for the background and 5 for the peak. This image is one of six images presented in Appendix~\ref{Sec:AppA}}.} 

         \label{Fig:image}
   \end{figure}

An image was reconstructed using the monochromatic version of the MiRA software of \cite{Thiebaut2008} \editguy{(\url{https://github.com/emmt/MiRA})}. 
As the (u,v) coverage was sparse, we used the same method as in \citet{Lacour2009} to regularize the image reconstruction process. The criterion minimized by MiRA \editguy{in this section} is a weighted sum of the squared distance to the data and of the squared distance to \editguy{an initial image (SD2I in Appendix~\ref{Sec:AppA}). The initial image} was built by fitting the data with a circularly symmetric star + shell model of \citet{Perrin2004}, the parameters of which are given in Table~\ref{tab:prior}. The total number of available data is 940 (705 squared visibilities and 235 closure phases). Images of 100x100 pixel were reconstructed. A hyper parameter $\mu$ controls the weight of the \editguy{regularization} in the reconstructed image with a total $\chi^2$ reading: $\chi^2_{\mathrm{tot}}=\chi^2_{\mathrm{data}}+\mu\chi^2_{\mathrm{rgl}}$. \editguy{$\chi^2_{\mathrm{data}}$ is defined and discussed in Appendix~\ref{Sec:AppA}.} The \editguy{higher} $\mu$ the higher the weight of the \editguy{regularization and its} impact on the reconstructed image. We tried several values of the hyper parameter and tested the impact on the image reconstructed with MiRA. \\
\editguy{We tried other reconstructions with MiRA or MACIM using various types of initial images and priors imposed by the regularization. They are discussed in Appendix~\ref{Sec:AppA} and the images are presented in Fig.~\ref{Fig:AppA}}. For the level of information available at the angular resolution obtained with IOTA, all images share the same basic features: 1) a bright central object of size $\simeq 20\,mas$ elongated in the north-south direction; 2) surrounded by a fainter environment of $\simeq 40\,mas$, roughly circular; 3) a dark patch in the environment \editguy{that is west} of the bright central object.
From among the available images, we selected the one with the lowest \editguy{hyper} parameter, $\mu = 10^8$, as it is more constrained by the data relative to the others. The image is shown in Fig.~\ref{Fig:image}.

\begin{figure}
   \centering
   \includegraphics[width=\columnwidth]{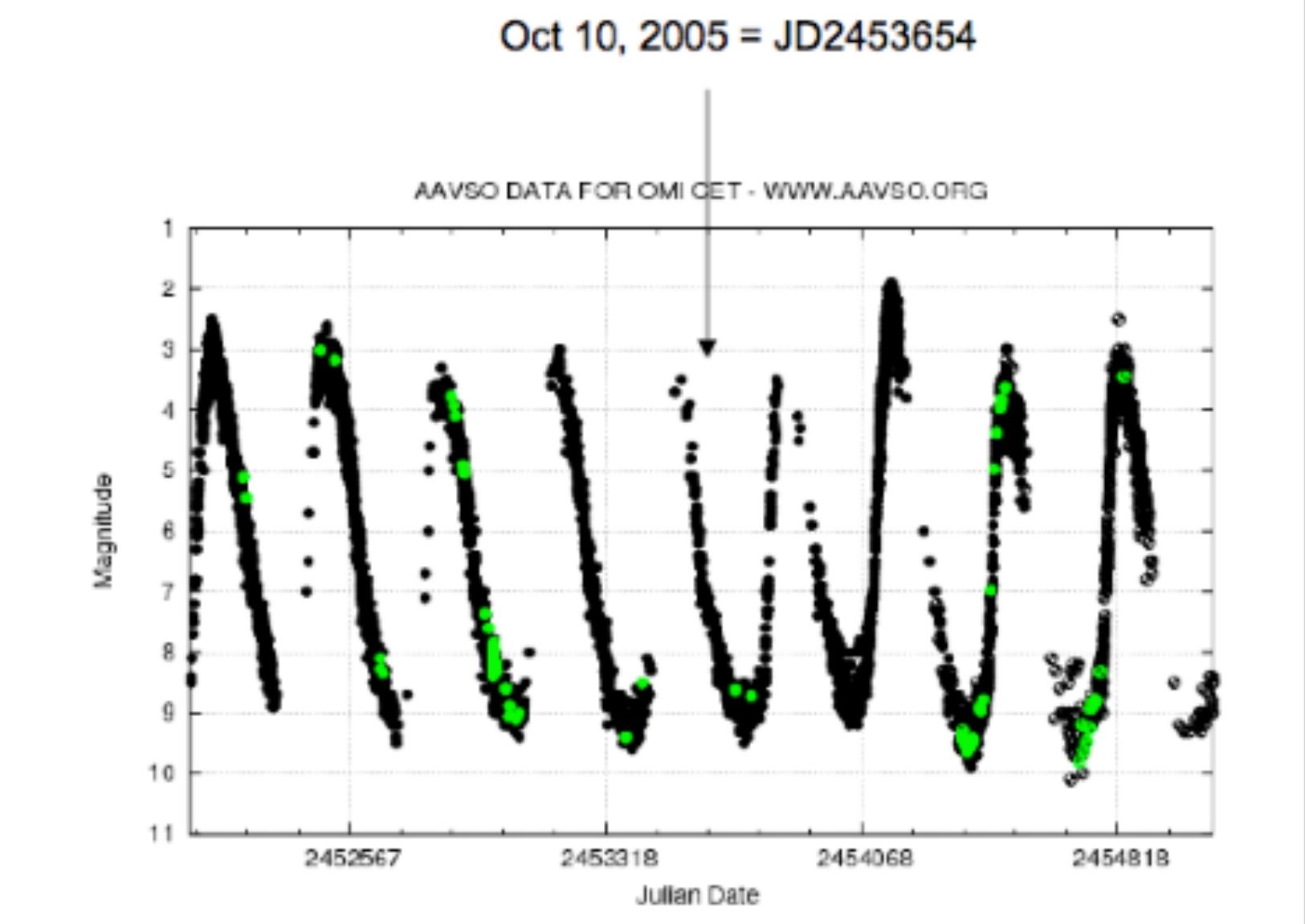}
      \caption{AAVSO V-band data for the period around the dates of the IOTA/IONIC observations.}
         \label{Fig:AAVSO}
   \end{figure}

\subsection{The inner envelope}

The image is generally consistent with the core-halo description of LPVs developed in \citet{Perrin2004}. The two-component, circularly symmetric model correlates well with infrared spectroscopy, recorded in the same spectral region \citep{Hinkle1984}.  The spectroscopy records molecular absorption against continuum emission from the effective photosphere. It is natural to associate the molecular absorption zone with the halo in Fig.~\ref{Fig:image} and to understand the core in the image as a representation of the deeper continuum-forming layer. This association is supported by the dynamical content of the spectra. The motions of the pulsating LPV atmosphere are followed in the \citet{Hinkle1984} study, showing in motions of gaseous CO the expansion, slowing, and falling back of the partially transparent halo.  The radial velocity of this layer with respect to the Mira CM velocity (as inferred from the centroid of low excitation envelope emission lines) can be used to follow the halo expansion through a pulsational cycle. Using an average envelope speed of 7.5\,km.s$^{-1}$ \citet{Hinkle1984} for half of the 321-day period, the total extent of the pulsational motion is found to be $1.1\times10^{8}$\,km. This corresponds well with the $1.4\times10^{8}$\,km radius difference between the top of the 10\,mas model halo and the photosphere (based on the Hipparcos distance of $2.8\times10^{15}$\,km). Thus, to good approximation, the inner boundary of the imaged and modeled halo shows the H-band photosphere and the outer boundary of the halo maps, approximately, the extent of the periodically pulsating gaseous layers. The minimum observed CO excitation temperatures, $\approx$2700K, are in between the inferred model layer temperatures of 2000K and 3400K.

What remains for us to consider is the relation with the \edit{extended molecular layers described by} \citet{Tsuji2000}. With a molecular excitation temperature of $\approx$ 1000K, it is, of course, outside the observed and modeled halo.  The coupling of the \edit{extended layers} to the star is not well observed. However,  \citet{hinkle2015} 
have shown in the LPV R\,Cas that gas at an intermediate temperature of 1300K shows a behavior that is intermediate with regard to the pulsating halo and the more stationary extended layers. The intermediate zone shows much-reduced mass motions, that are not coherent with the halo pulsation, and are consistent with "puffs" of mass loss, maintaining line-of-sight coherence over two pulsation cycles, without yet attaining escape velocity and propelling mass loss.  While longer spectroscopic time series of additional targets are needed to confirm this finding, it is appears to be a direct observation of the process of populating the \edit{extended layers} with material raised above the pulsating halo.  Expecting a comparable phenomenon in the similar LPV Mira, we suggest that the evidence favors dust formation, or at least critical grain growth, which is likely to occur in either the halo or the \edit{extended layers}, where a combination of low ambient temperature and relatively enhanced density favors dust formation.

A core-halo structure similar to Fig.~\ref{Fig:image} has been reported in interferometric imaging of the LPV T Lep \citep{Le_Bouquin2009}.
An important difference is that the T Lep star is approximately circularly symmetric.  The Mira image is not symmetric. In the following, we consider what the asymmetry may have to tell us about the envelope structure and, possibly, about the mass-loss process.


\subsection{Interpretation of the image}
The bright core identified with the photosphere shows an elongated structure.  A number of studies have reported strong azimuth dependence in the apparent size of LPV stars. Various models have been proposed, ranging from underlying non-radial oscillation to less dramatic but equally puzzling phenomena such as an azimuthally dependent envelope shape. While non-radial oscillation has not been ruled out for Mira by the new image, the strong asymmetry that is a primary feature of the observation indicates a more complex appearance than a simple elliptical shape of the photosphere or envelope. This is confirmed in the next section with the technical argument that the data are hard to fit solely with  an elongated structure and, in fact, an asymmetrical brightness is required.  \\

A natural question is to ask whether this differential effect represents a dimming or a brightening.  This can be answered with reasonable confidence by referring to the AAVSO light curve for the last 15 years reproduced in Fig~\ref{Fig:AAVSO}.  This shows that the 2005 maximum of Mira was one of the faintest in recent years and more than a magnitude fainter than the second following maximum. At the same time, the image appears to show that most of the stellar disk is rather uniformly bright. We tentatively identify the asymmetry with a differential darkening phenomenon.  

A related question considers whether the dimming represents a zone of reduced brightness on the photosphere or a region of increased opacity in the envelope. If the dimming is due to a dark region on the star, a possible explanation would be a long-lived convective structure.  Convection is expected to be important in LPVs and present at the same time as pulsation. We note that pulsation does not transport energy effectively and convection is still required for energy transport \citep{wood2007}.  The relative apparent brightness differential across the photosphere (approximately $6\times$), which implies a similar differential in total surface emittance, also seems too large in the H band for a temperature differential.  The dark region covers a significant fraction of the hemisphere and even though large convective cells are predicted, this scale seems unprecedented for a convective effect. The 3D hydrodynamical models of AGB stars combine the effects of pulsation and convection, providing evidence that convection plays a role to produce surface brightness asymmetries \citep{Freytag2017}. These simulations show that dark patches cannot be present alone without bright patches. The spatial convective signature is better characterized as localized, high-contrast bright spots associated with hot, rising material, rather than localized dark spots of falling, cool material. Since the present data set do not demonstrate or require a bright patch, we take this as an additional confirmation for a dark region.\\

The darkening could be due to an increased opacity in the halo or above. This appears more consistent with the apparent large spatial scale - the dark spot appears to significantly obscure a region of the halo.  It is also consistent with previous evidence for irregularities in the near circumstellar region of LPVs, including Mira \citep{lopez1997}.
Therefore, we suggest that the Mira image is showing the effects of a localized enhancement in opacity above the halo. This could be consistent with localization within the intermediate layer feeding the \edit{extended molecular layers} or with \edit{that material itself}, with coherent structure lasting over several pulsational periods, but not indefinitely consistent with the observed shell expansion velocity $\approx$few km/sec - \citet{hinkle2015}. It is reasonable to find that the cause is absorption by dust located \edit{in or near the extended molecular layers} as the dust condensation radius is expected in this region, as discussed in, for example, Reid \& Menten 1997, Perrin et al. 2004 \& 2015, and Wittkowski et al. 2007. In the next section, an estimate of the mass of the clump with respect to the net mass loss rate leads to another plausible explanation for a location near or somewhat interior to \edit{this region, which \citet{Tsuji2000} describe as a MOLsphere.} \\

Previous images of LPVs with large telescope aperture masking \citep{Woodruff2008} show a rich variety of information about the upper layers of LPVs, but they do not probe asymmetries in the apparent shape of the photosphere well owing to the lack of (u,v) sampling beyond the first visibility null.  Multi-telescope interferometry with COAST has reported some significant non-zero closure phases, but did not distinguish the spatial location of the responsible asymmetries \citep{Burns1998}.  Hence the inference that asymmetry and inhomogeneity in the circumstellar dust opacity is consistent with previous imaging work, but not well constrained by it. The objective of the next section is to study what constraints on the opacity distribution are provided by the Mira interferometric image.


\section{Obscuring dust patch}
\label{Sec:model}

\begin{figure*}
\begin{tabular}{cc}

     \vspace{-0.24cm} 
     \includegraphics[width=\columnwidth]{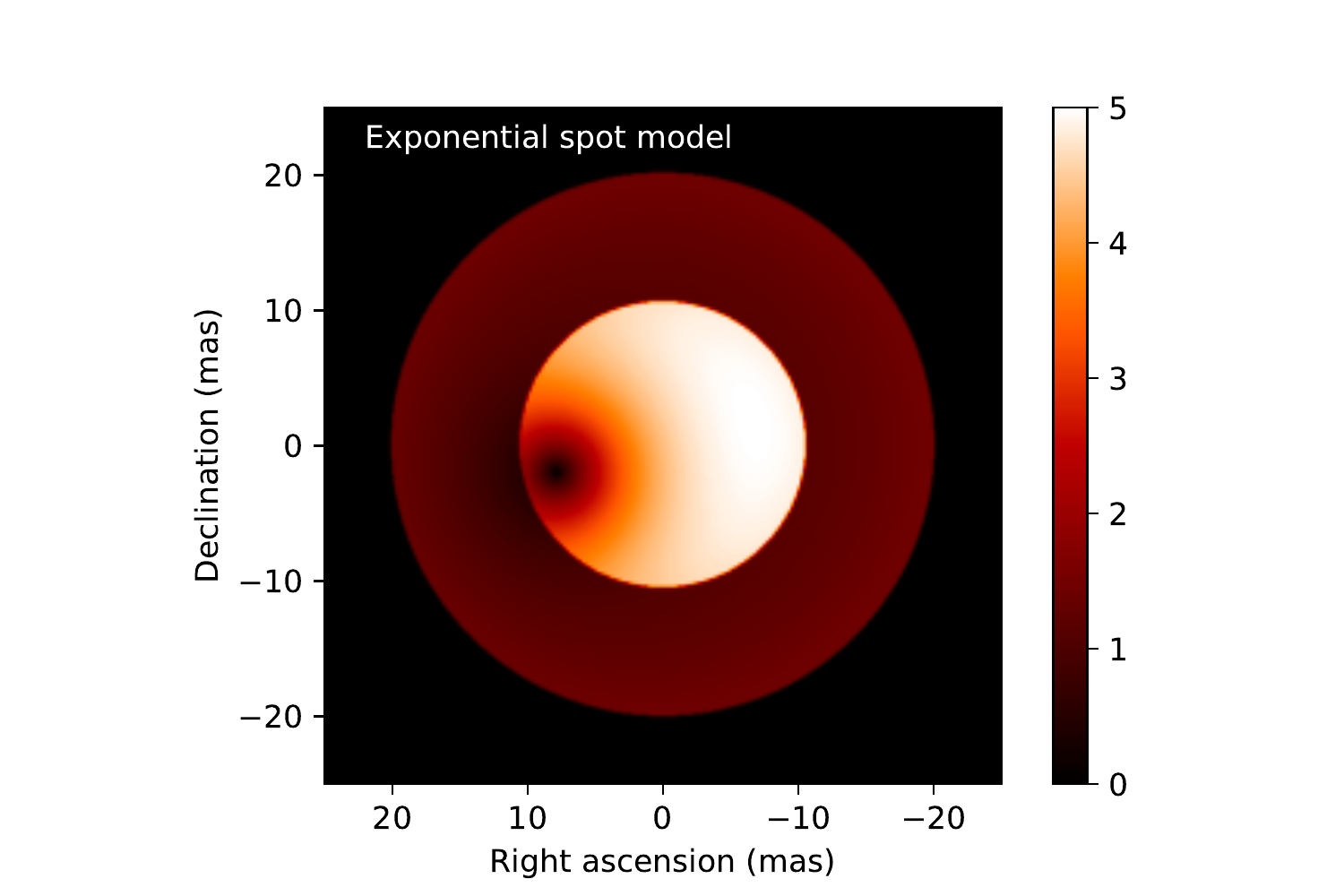} & \includegraphics[width=\columnwidth]{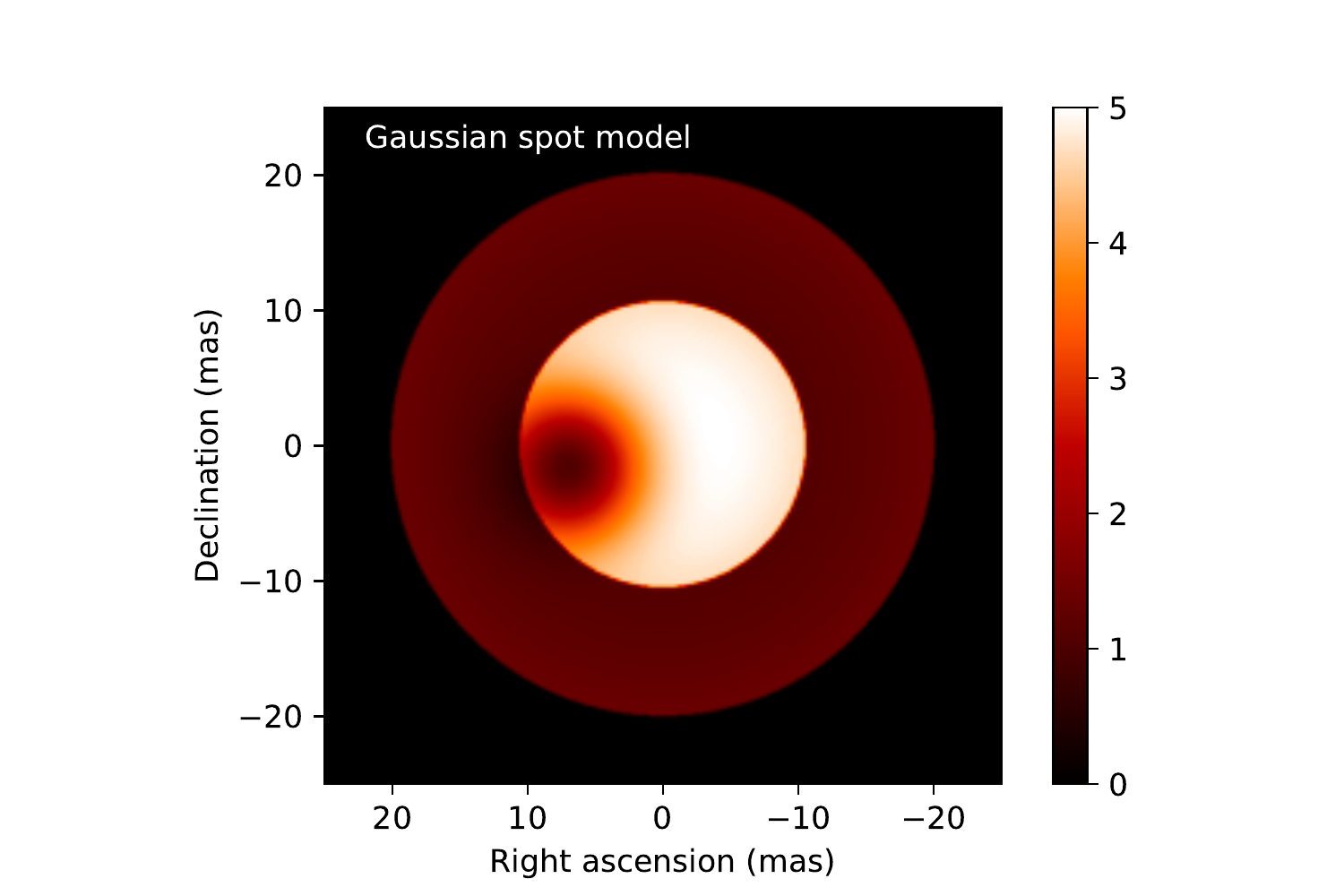} \\
      
       
    \includegraphics[width=\columnwidth]{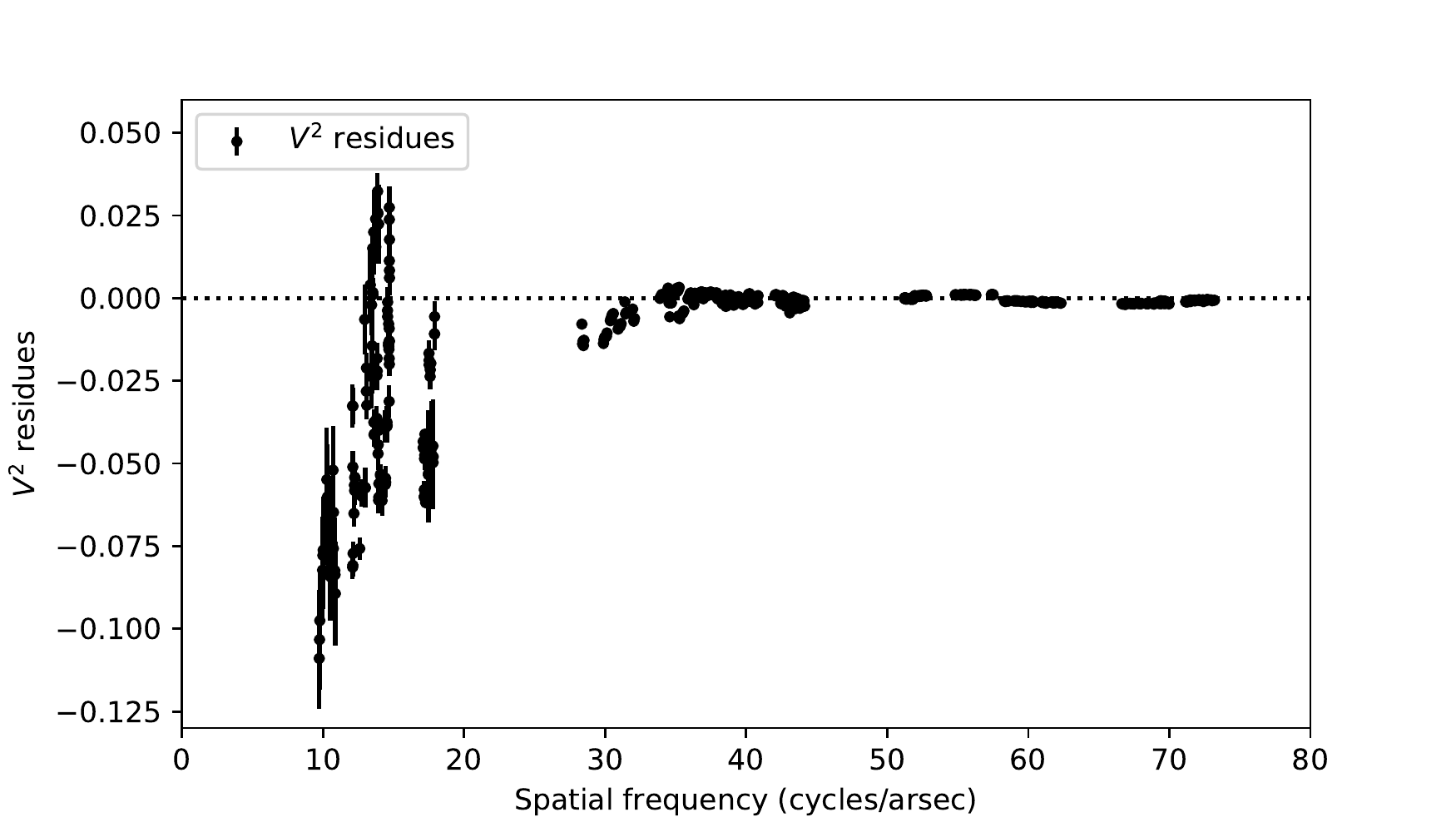} & \includegraphics[width=\columnwidth]{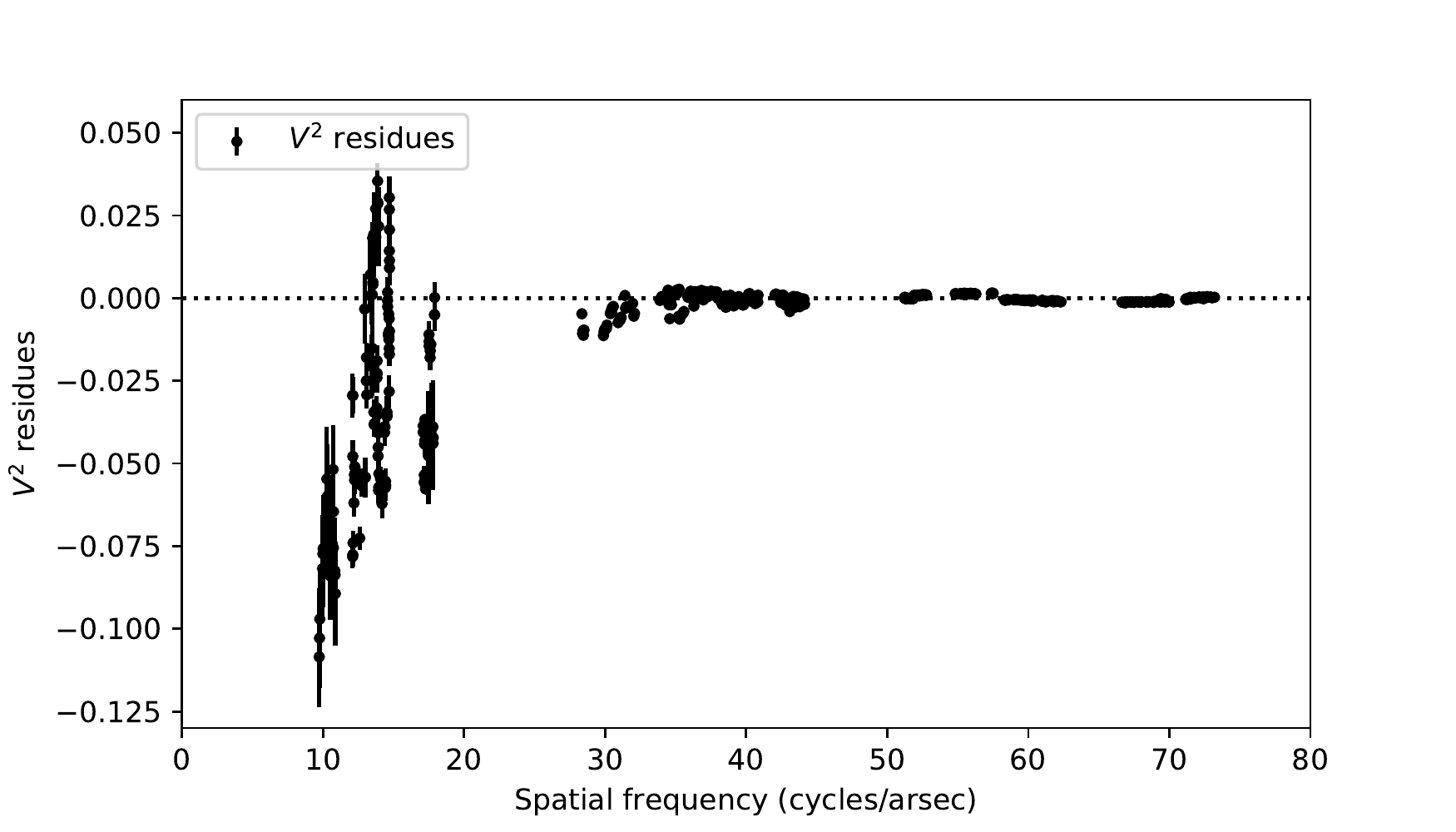} \\
    \includegraphics[width=\columnwidth]{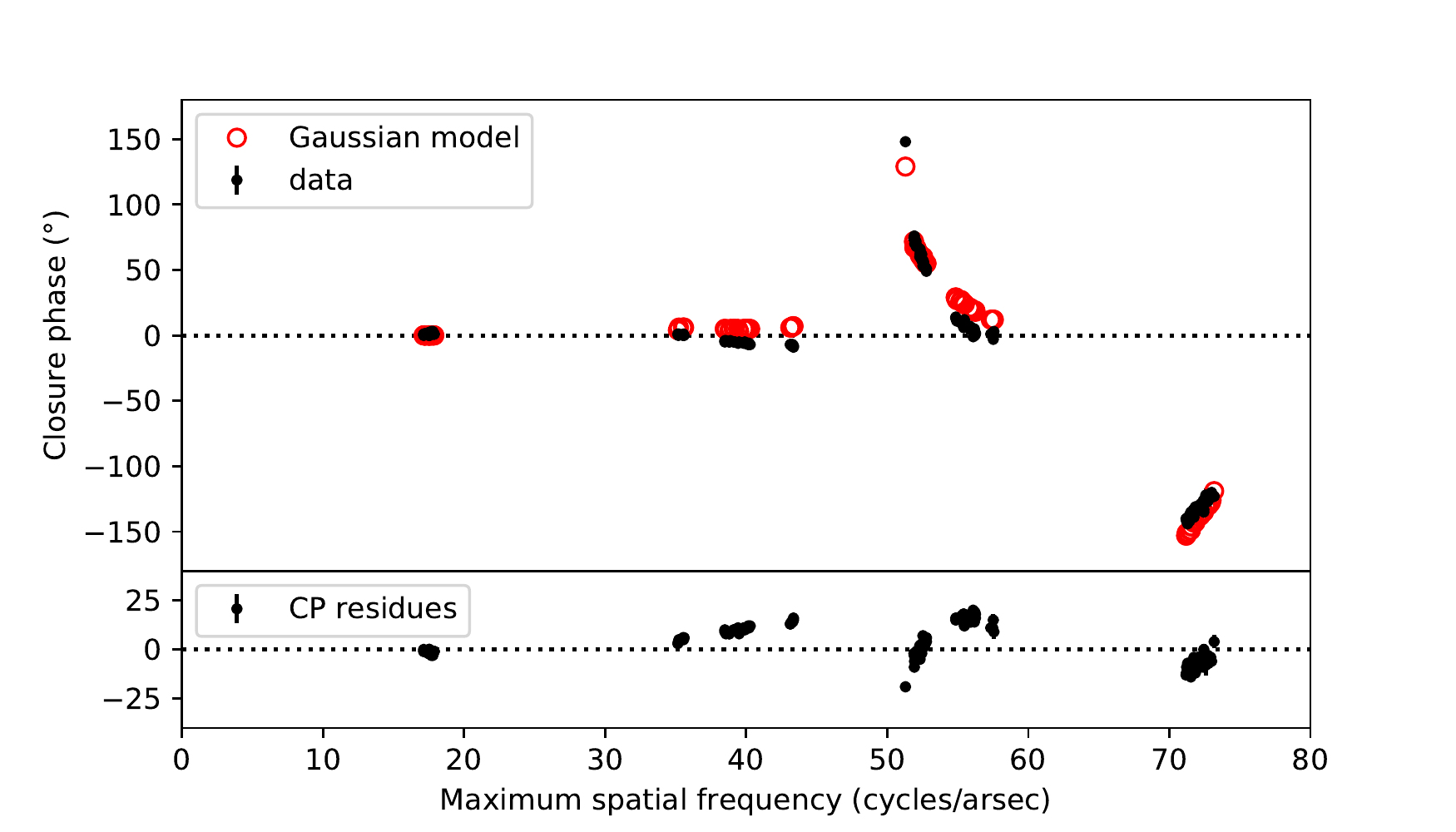} & \includegraphics[width=\columnwidth]{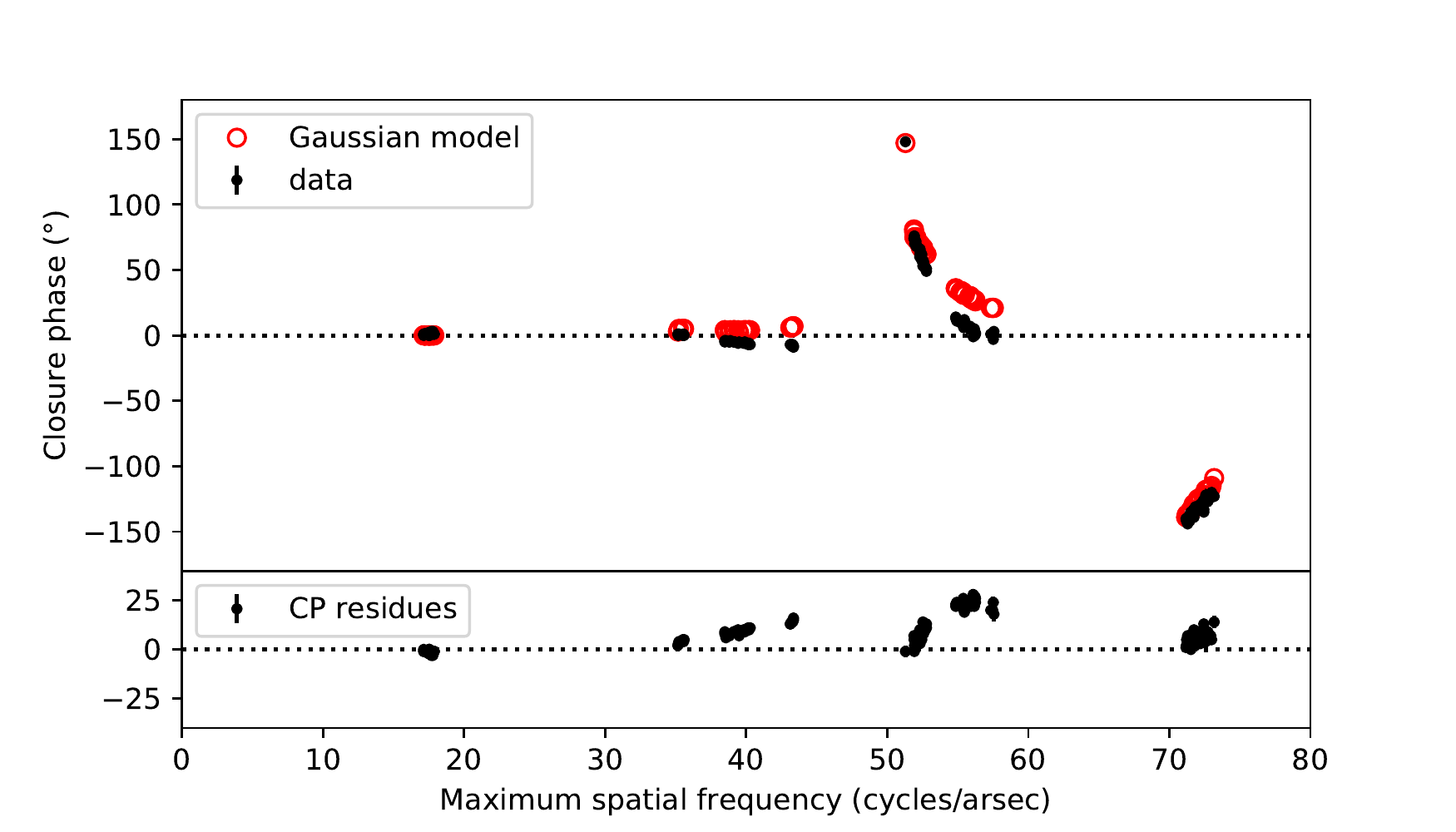} \\
\end{tabular}
\caption{Result of the fits with the patchy model. Left: Exponential attenuation model. Right: Gaussian attenuation model. From top to bottom: Surface brightness distribution, $V^2$ residues, closure phases (red), model closure phases (black), and closure phases residues. \editguy{The closure phase and closure phase residues are displayed as a function of the maximum spatial frequency of the triangle of baselines.}}
\label{Fig:patch_fit}
\end{figure*}

\subsection{Dust patch model}
\label{Sec:Dust_patch_model}
The image in Fig.~\ref{Fig:image} clearly shows asymmetries that require a more complex model than the one used in Sect.~\ref{Sec:prior}. We have tried several approaches to explain the data with a central source elongated in the north-south direction and various offsets with respect to the more circularly symmetric shell. These did not lead to satisfactory fits as, in particular, they were inconsistent with the closure phases. \\

Looking at the reconstructed image, however, it appears generally compatible with the simple circularly symmetric model of Sect.~\ref{Sec:prior} in the sense that it is globally symmetric at low spatial frequency scale and with characteristic sizes of 20 and 40\,mas for the star and shell around it. With the model temperatures fixed as described above, the inner and outer diameters and the optical depth of the shell are essentially given by the first visibility lobe and the zero crossing around 50\,cycles/arcsec. The parameters of the star+shell model are therefore those of Table~\ref{tab:prior}. The star+shell model has five free parameters. It defines a featureless surface brightness distribution $B_0(x,y)$. After determining the minimal impact of adding high spatial resolution features to the star+shell model, we decided to separately fit  the parameters of Table~\ref{tab:prior} and the parameters of the more complex model described below. This method with two separate steps led to satisfactory results.\\

To fit the closure phases and account for the azimuth dependency of the squared visibilities, we considered a dark patch to the \editguy{west} combined with the star+shell model.  We tried a single patch as increasing the number of the patches made the fit more complex but did not bring improvement.

Experimenting with several patch options, a smooth edge better matches the data. We used two types of smoothing functions: exponential and Gaussian. Technically, the star+shell model is multiplied by an attenuation function:

\begin{equation}
    f(x+\alpha,y+\delta)= 
\begin{cases}
    1-a.e^{-\frac{2 \ln 2 \sqrt{x^2+y^2}}{\mathrm{FWHM}}}, \mathrm{\;\;if\;exponential\;patch} \\
    1-a.e^{-\frac{4 \ln 2 \left(x^2+y^2 \right)}{{\mathrm{FWHM}}^2}}, \mathrm{\;\;if\;Gaussian\;patch}
\end{cases}
\label{Eq:attenuation}
.\end{equation}

The dark patch model then has four parameters: 1) the patch offset $\alpha$ and $\delta$ with respect to the star center; 2) the attenuation strength $a$ (between 0 and 1); 3) the patch FWHM. The total number of parameters of the star+shell+patch model is 9 and the corresponding surface brightness distribution is $B(x,y)$ with $B(x,y)=B_{0}(x,y)\times f(x,y)$. \\


The closure phases outside the $\pm 150\degree$ range were found to have a diverging impact on the fit as they are very sensitive to noise. We therefore decided to crop them to search for the most robust solution. The best parameters we found are listed in Table~\ref{tab:patch_model}. The error bars were derived by varying the $\chi^2$ by 1. It is clear, however, that these are lower limits since the model does not perfectly account for the data. We did not consider it useful to refine error bars as the goal is to obtain orders of magnitude for physical parameters of the dust clump later in Sect.~\ref{Sec:dust} and as we used two different models for the patch which give an idea of the scatter for these parameters as reported in Table~\ref{tab:attenuation}. \\


The surface brightness distributions, residues, and closure phase fits are presented in Fig.~\ref{Fig:patch_fit}. The fit to the closure phases is quite good, even at high spatial frequencies, with residuals smaller than $25\degree$. The shape of the closure phases is consistent with a smooth spot as a uniform spot did not provide such nice fit. The $V^2$ residuals at high visibility are quite large. We investigated this issue by fitting the low spatial frequency data (below 40 cycles/arcsec) with a uniform disk model with an azimuth-dependent diameter ($10\degree$ bins). The amplitude of the scatter remained the same albeit more symmetric around 0. We concluded that the scatter is real and does not depend on the model we chose to interpret the data. \\

 The two models yield qualitatively similar results and although the $\chi^2$ for the Gaussian model is a bit lower, both were retained for further consideration. The total $\chi^2$ may look large but the two branches of models capture most of the visibility and closure phase features that are in very good agreement with the reconstructed image so that we consider that it is a satisfactory description of the object.

\begin{table}
\caption{Best parameters for the dark patch model fit of the data with an exponential and a Gaussian attenuation function.}
\centering
\begin{tabular}{ccc}
\hline \hline
& \multicolumn{2}{c}{Dark patch model}\\
\cline{2-3}
& Gaussian & Exponential \\ 
\hline
FWHM (mas) & $8.57\pm0.40$ & $6.89\pm0.65$ \\
$\alpha$ (mas) & $6.94\pm0.23$ & $7.84\pm0.30$ \\
$\delta$ (mas) & $-1.69\pm0.15$ & $-2.04\pm0.21$ \\
$a$   & $0.783\pm0.005$  & $1.00^{+0.00}_{-0.008}$   \\
$\chi^2_{V^2}$ & 134 & 165 \\
$\chi^2_{\mathrm{CP}}$ & 83 & 69 \\
$\chi^2_{\mathrm{tot}}$ & 124 & 146\\
\hline
\end{tabular}
\label{tab:patch_model}
\end{table}

\subsection{Interpretation of the attenuation function}
We concluded that the dark patch is probably located above the pulsation/convective zone and is accounted for by absorption from a patch of material. It therefore equally absorbs the radiation from the star and from the inner envelope. We neglect the H-band emission by the patch as it is at a higher, cooler elevation.  The attenuation function can be locally equated to an optical depth effect:
\begin{equation}
    f(x,y)=e^{-\tau(x,y)}
.\end{equation}
The surface brightness distribution then is written as:
\begin{equation}
    B(x,y)=B_{0}(x,y) \times  e^{-\tau(x,y)}   
.\end{equation}

We can therefore derive the local optical depth from the ratio of the obscured to unobscured surface brightnesses. The local optical depth can then be converted into a mass of absorbing material. Let $\rho$ [g.cm$^{-3}$] be the density of the absorbing material, $\kappa$ [g$^{-1}$.cm$^2$] the opacity and $z(x,y)$ the local depth of absorbing material. In the following relation:
\begin{equation}
    \frac{\tau(x,y)}{\kappa}=\rho z(x,y),
\end{equation}
where $\rho z(x,y)$ is the mass column density in g.cm$^{-2}$. The integration of this quantity over the stellar disk, therefore, yields the total mass $m$ of absorbing material, $m$:

\begin{equation}
    m = \iint{\rho z(x,y) \, dxdy}
.\end{equation}

The integral can be evaluating using the surface brightnesses of the star+shell and star+shell+patch models:

\begin{equation}
    m = \frac{1}{\kappa}\iint{-\ln \left[ \frac{B(x,y)}{B_{0}(x,y)} \right] dxdy}
.\end{equation}

We can also evaluate the average attenuation by the patch with the ratio of the integrated surface brightnesses:

\begin{equation}
    \bar{f}=\frac{\iint{B(x,y) \, dxdy}}
           {\iint{B_{0}(x,y) \, dxdy}}
.\end{equation}

The results are given in Table~\ref{tab:attenuation}. Uncertainties were derived with the Monte-Carlo technique using the values of Table~\ref{tab:patch_model}. The two patch models yield similar results. The patch causes an average attenuation of the H-band flux of $0.925\pm0.007$ (minimum relative surface brightnesses of $2\times10^{-3}$ and $2\times10^{-1}$ for the exponential and Gaussian respectively, compatible with the dynamic range of the reconstructed image). We use the average value of $(6.31\pm0.72)\times10^{25}\,$cm$^2$ for the mass-opacity product (optical depth integrated over the star surface), denoted as $m.\kappa_{\mathrm{H}}$ from this point forward.

\begin{table}
\caption{Comparison of the average attenuations $\bar{f}$ and integrated optical depths $m.\kappa$ for the two models of patches.}
    \centering
    \begin{tabular}{lcc}
    \hline \hline
    Patch model    & Exponential & Gaussian \\
    \hline
    $\bar{f}$  &   $0.927\pm0.007$   & $0.924\pm0.005$ \\
    $m.\kappa$ (cm$^2$) & $(6.05\pm0.72)\times10^{25}$  & $(6.57\pm0.67)\times10^{25}$ \\
    \hline
    \end{tabular}
     \label{tab:attenuation}
\end{table}

\begin{table*}
\caption{Dust parameters and mass loss rates for silicate species for small or large grains.}
\centering                          
\begin{tabular}{clll}        
\hline\hline                 
Particle size range & Opacity range & Dust mass & Gas mass \\
& $\kappa_{\mathrm{H}}$(cm$^2$g$^{-1}$) & (M$_{\odot}$) & (M$_{\odot}$) \\

\hline                        

1 -- 5.5 nm & 0.6 -- 1.45 & 2.2 -- 5.3 $\times 10^{-8}$ &  2.0 -- 4.7 $\times 10^{-6}$\\
0.45 -- 0.47 $\mu$m & 1.07 -- 1.27$\times 10^6$ & 2.5 -- 3.0$\times 10^{-14}$ & 2.2 -- 2.7 $\times 10^{-12}$ \\
\hline                                   
\end{tabular}
\label{table:dustmass}      
\end{table*}

\subsection{Interpretation of the attenuation as a dust opacity effect}
\label{Sec:dust}
As discussed in Sect.~\ref{Sec:imaging}, our favored hypothesis to explain the lower surface brightness at the spot is dimming by a dust patch. Al$_2$O$_3$ is a good candidate for dust formation at high temperature in the molecular shell of Mira stars (see e.g. \citet{Perrin2015}). But the condensation of Al$_2$O$_3$ at  distances and concentrations implied by observations requires high transparency of the grains in the visual and near-IR region to avoid destruction by radiative heating \citep{Hofner2016}. We therefore consider silicate grains to explain the obscuration. Amorphous Mg$_2$SiO$_4$ grains are the main candidate for driving the winds of O-rich AGBs according to \citet{Bladh2012}. We interpret the dimming in H measured with IOTA/IONIC and in V from the AAVSO as being due to this single dust component. Within the scope of the single-grain type hypothesis, the relative V/H extinction constrains the dust grain characteristics. We have decided to consider two types of magnesium-silicate dust grains: MgSiO$_3$ and Mg$_2$SiO$_4$. As a remark, we note that \citet{Khouri2018} did their modeling with aluminum oxides, adding that magnesium-silicate dust grains have scattering properties that are similar to those of Al$_2$O$_3$.\\

Opacities can be obtained from the literature for specific grain sizes and types. We performed calculations of the opacity of dust in the Mie approximation using the Fortran code, {\it bhmie,} from \citet{Bohren&Huffman1983}, which was modified by Bruce Draine (http://scatterlib.wikidot.com/mie).  Optical constants were taken from \citet{Jaeger2003}. The numerical values are listed at the Jena group web page: http://www.astro.uni-jena.de/Laboratory/OCDB/amsilicates.html (OCDB database). \\

\begin{figure*}
\centering
\begin{tabular}{cc}
 \includegraphics[width=0.95\columnwidth]{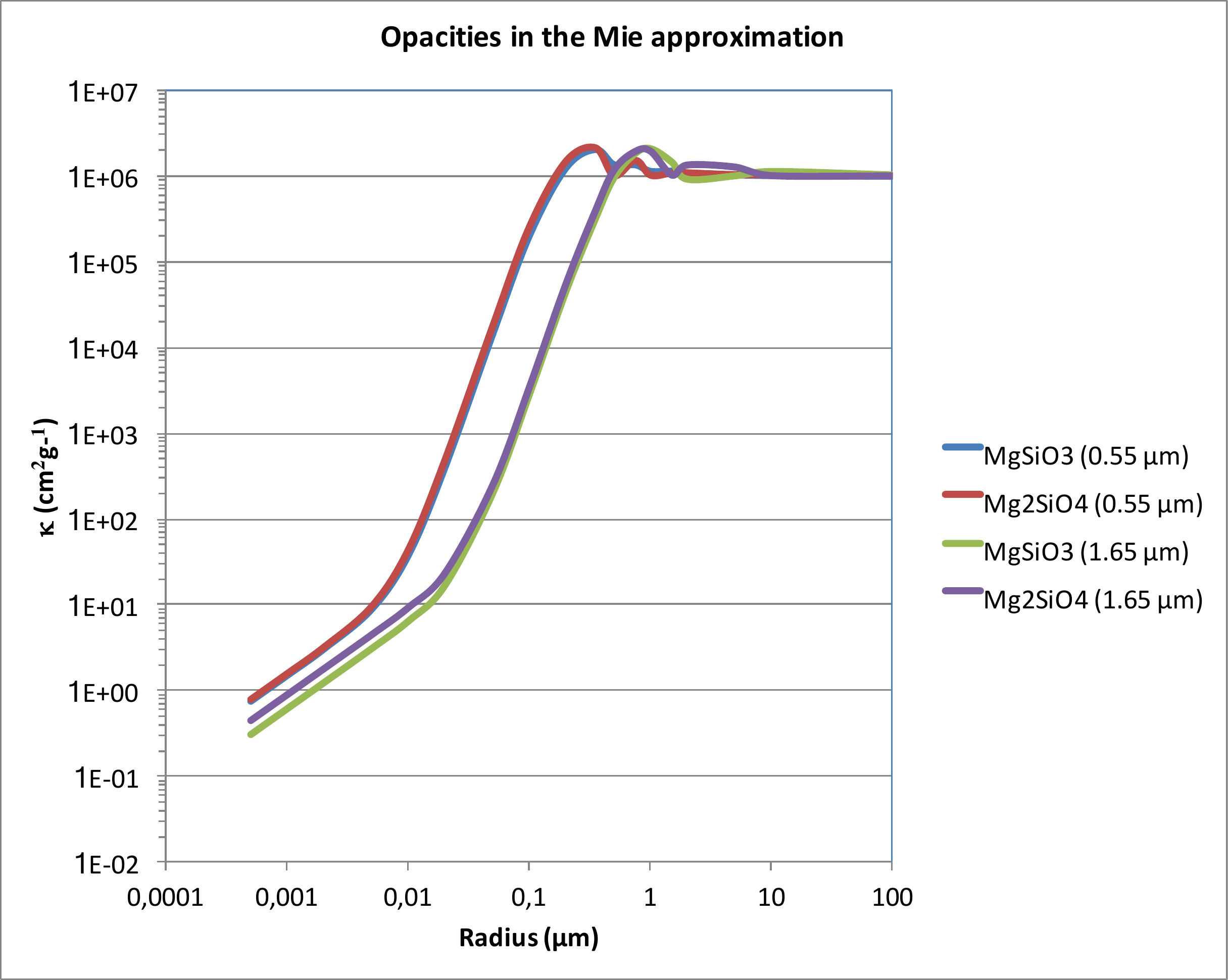} & \includegraphics[width=0.95\columnwidth]{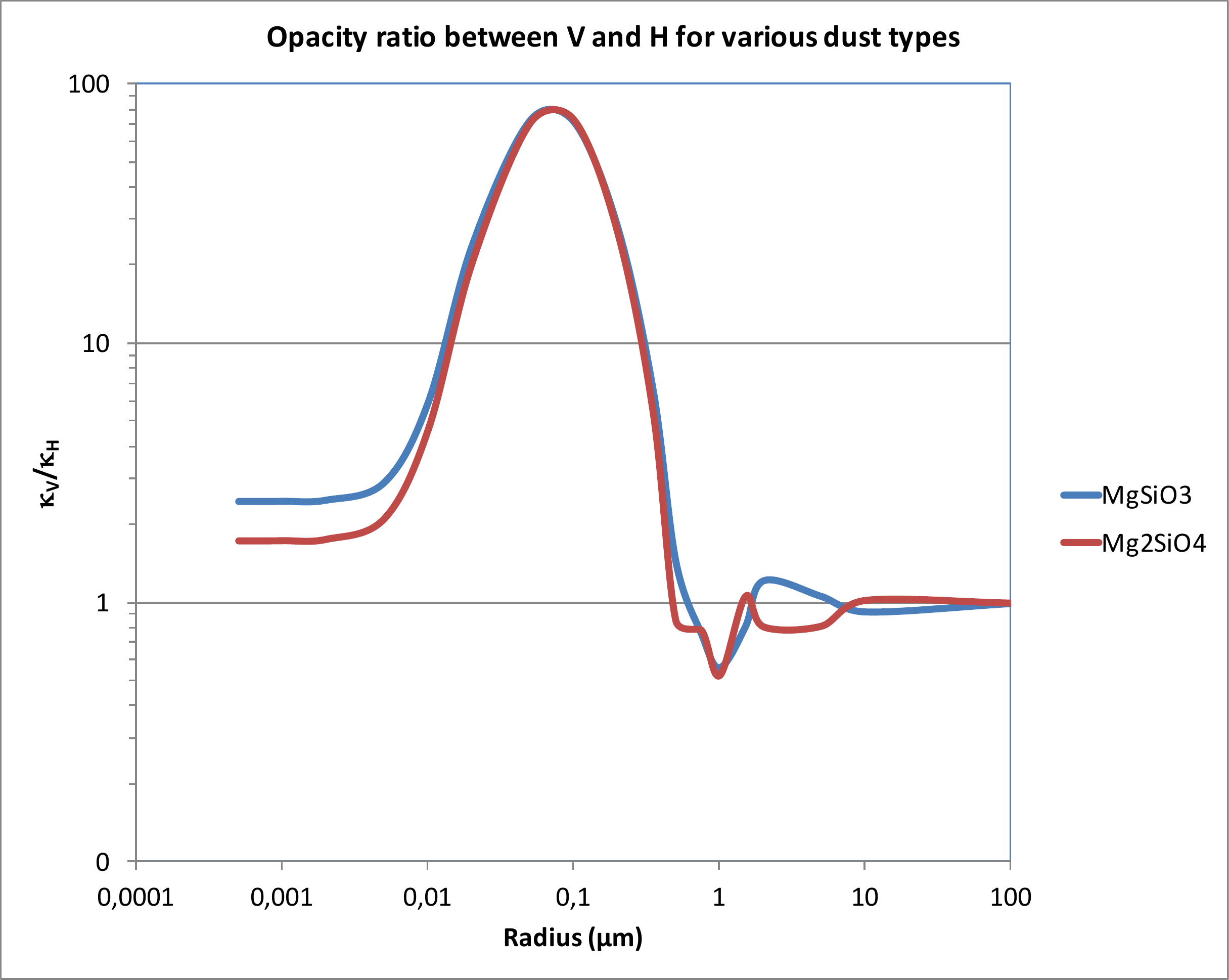} \\
\end{tabular}
\caption{Result of computations done with the {\it bhmie} code for MgSiO$_3$ and Mg$_2$SiO$_4$ dust with grain sizes ranging from 0.5\,nm and 100\,$\mu$m. Left: Opacities at V and H. Right: $\kappa_V/\kappa_H$ ratio.}
\label{Fig:Mie}
\end{figure*}

The extinction efficiency $Q_{\mathrm{ext}}$ was computed at 0.55 and 1.65\,$\mu$m wavelengths.  In the Mie approximation, the extinction efficiency can be translated to an opacity via the equation:
\begin{equation}
 \kappa=\frac{3Q_{\mathrm{ext}}}{4a\rho}  
,\end{equation}

where $a$ is the radius of the dust sphere and $\rho$ is the density, all in CGS units. The typical density of 3 g.cm$^{-3}$ from the OCDB database was used. The opacities were computed for radii ranging from 0.5\,nm to 100\,$\mu$m to explore the different regimes of the Mie scattering. The plots of Fig~\ref{Fig:Mie} show the opacities for the two species of dust as a function of radius at 0.55 and 1.65\,$\mu$m as well as the $\kappa_{\mathrm{V}}/\kappa_{\mathrm{H}}$ ratio. The opacities are quite similar for the two species of dust except for radii below 0.01\,$\mu$m. \\

The AAVSO data of  Fig~\ref{Fig:AAVSO} show that the two V-band maxima before and after the median date of the observation (JD2453654) are brighter, respectively, by a 0.4 and 0.0 magnitude. Assuming the difference in maximum magnitude is entirely due to the obscuration of a persisting dust cloud, the V-band data give an upper limit on the attenuation due to dust of 0.4 magnitude. We made the following assumptions to compute the flux attenuation by the dust patch in V: 1) the stellar brightness in the visible is dominated by the brightness of the molecular shell, the photosphere is not visible: this is supported by the large optical depth of the shell at visible wavelengths derived both from observations (diameters in the visible are much larger, see for example the first observational evidence by \citet{Labeyrie1977}) and by the 3D hydrodynamical simulations; 2) the diameter of the molecular shell is the same as in H; 3) the dark patch characteristics are the same as at H except for the optical depth. 

With these assumptions, the maximum global flux attenuation can be computed as a function of the strength parameter $a$  in Eq.~\ref{Eq:attenuation}. The maximum attenuation is 5\% with the Gaussian model, whereas it reaches 10\% with the exponential model. In consequence, we used the exponential model to explore the variations of the attenuation and of the $m.\kappa_{\mathrm{V}}$ product versus {\it a}. The maximum for the geometrical characteristics of the patch derived in the H band is an attenuation of 10\% of the flux corresponding to a difference in magnitude of $\Delta V = 0.11$. This is compatible with the maximum brightness decrease with respect to the two neighboring V-band maxima. This cannot, however, be the only source of the dimming in V as the AAVSO data show cycle-to-cycle variations up to a magnitude or more. Unless several such patches are episodically produced. Here, we acknowledge the simplicity of our approach as the reasons for the cycle-to-cycle V magnitude variations of Mira are likely to be multiple and not strictly due to dust patches occurring at the surface of the star. Our simple hypothesis, therefore, provides an upper value for the opacity of the dust patch in V. Following this approach, it yields a maximum mass-opacity product at visual wavelengths of $m.\kappa_{\mathrm{V}}=1.49\,10^{26}\,$cm$^2$ and, thus, a maximum ratio for the dust opacity between V and H of:
\begin{equation}
    \frac{\kappa_{\mathrm{V}}}{\kappa_{\mathrm{H}}}\leq\frac{1.49\times10^{26}}{(6.31\pm0.72)\times10^{25}} \simeq 2.36\pm0.27
.\end{equation}

The relative fraction of V/H extinction assumed does not strongly impact the deductions concerning dust mass.   In any event, the inferred V-band obscuration cannot explain the large variations of the maximum visual magnitude shown by the AAVSO measurements. 


According to Fig.~\ref{Fig:Mie}, this is compatible with both MgSiO$_3$ and Mg$_2$SiO$_4$ dust. The particle radii ranges are shown in Table~\ref{table:dustmass}, along with the implied total mass of the dust in the patch, and also the total mass of the associated gas - assuming a gas/dust ratio of 89.5 \citep{knapp1985}.



\section{Discussion}
\label{Sec:discussion}

In the previous sections, we modeled the IOTA/IONIC data on Mira in the H band using a patch of absorbing material obscuring the core-halo, that is, the modeled H-band photosphere and pulsating envelope. We show, based on a combination of these data with visual photometric data, that this material is compatible with silicate dust. As summarized in Table~\ref{table:dustmass}, the inferred dust mass associated with the absorbing material is strongly dependent on the dust physical characteristics.  Our data set is limited and the characteristics of the dust remain rather speculative in this analysis and further interpretation must rely on indirect evidence. 

Since micron size grains are commonly inferred from models of the dust at several AU from an LPV \citep{Ohnaka2016}, such a grain size might be favored for the observed patch.  However, from Table~\ref{table:dustmass}, the associated dust/gas mass would be only $\approx 10^{-12}$\,M$_{\odot}$.  This is much less than the known yearly mass loss rate for Mira of $3\times10^{-7}\,$M$_{\odot}/yr$ \citep{Ryde2000}. In order for such a patch to contribute significantly to the observed mass loss rate, it would be necessary to suggest an implausibly large number of patches or high frequency and speed of ejection. So we tentatively reject the large grain option for the observed patch.

\edit{There is a precedent for considering small grains. Recent studies of the mira W Hya by \citet{Ohnaka2017} show a shift toward smaller grain sizes near photometric minimum (0.54), compared to their earlier study at 0.77 (\cite{Ohnaka2016}). This is not surprising since the minimum brightness corresponds approximately to maximum expansion and greatest atmospheric cooling. Dynamical models reported by \citet{Hofner2016}, \citet{Aronson2017} and \citet{Liljegren2017} indeed show dust formation near the brightness minimum. The observations reported here are also near minimum.}

From Table~\ref{table:dustmass}, a small grain size would be consistent with a reservoir of gas/dust mass in the patch  $\approx10$ times larger than the annual stellar mass loss if the gas/dust ratio can be applied at short distance from the photosphere and comparable to or a bit below the yearly mass loss rate otherwise.
Therefore, it is reasonable to consider the possibility that the patch can be linked with the mass ejection process and what this implies for the ejection efficiency. 


Only a single patch was observed. However, from the model suggesting that the patch is near but outside the halo and from the knowledge that approximately 1/3 of the progeny of LPVs show an asymmetry \citep[for Mira, here, attributed to dust patches]{Bobrowsky2007}, it can be estimated from geometry and assumed random distribution that at any given time there may be on the order of five such patches around Mira, of which zero or one would typically be in the line of sight to the halo at any given time.  Thus, the elevation of a few percent of each of the mass of each of these "puffs" of material over each pulsation cycle, on average, could support the observed Mira mass loss rate. \citet{Chandler2007} reported their detection of episodic mass loss around Mira at 11.15\,$\mu$m with the ISI interferometer. Our detection in H band is compatible with their finding.

With low envelope expansion velocities, compared to the escape velocity, and being on the same order as those observed \edit{in the extended molecular layers}, 
puffs of rising dust-enhanced material would have ample time for grain growth during the elevation to the vicinity of 2+ stellar radii, where much larger grain sizes (with much reduced opacity) are found by modeling spatially resolved polarimetric observations: \citet{Norris2012} have detected 0.3\,$\mu$m grains of Mg$_2$SiO$_4$ or MgSiO$_3$ for W Hya, R Dor and R Leo at 2 stellar radii or less; \citet{Ohnaka2016} detected 0.4-0.5\,$\mu$m  grains of Al$_2$O$_3$, Mg$_2$SiO$_4$ or MgSiO$_3$ between $\approx$2-3 stellar radii in clumpy dust clouds for W Hya; \citet{Adam2019} detected 0.1\,$\mu$m grains of Al$_2$O$_3$, Mg$_2$SiO$_4$ , or MgSiO$_3$ in a clumpy shell around IK Tau between 2 and 5 stellar radii; \citet{Khouri2016} measured the asymmetric photosphere of R Dor attributed to scattering by dust. Finally, \citet{Khouri2018}  observed clumpy structures in visible polarized light around Mira at larger distance than our measurements and within $\approx$8 photospheric radii. They used aluminum oxide grains to model their data and concluded that aluminum could be in dust phase if the grain size is less than 0.02\,$\mu$m. All other existing observations, apart from our own and those of \citet{Khouri2018}, detect a few clouds of larger dust particles that account for a good fraction of the mass loss. This is a discrepancy that is to be resolved with more measurements. The greater grain size at higher elevation satisfies the preferred conditions to drive mass loss by radiation pressure on the grains \citep{Hofner2008}, that is, 0.36-0.66\,$\mu$m at 2-3\,$R_{\star}$.


These findings give rise to a few more questions. The first two regard the nature of the mechanism that could explain the existence of such a clump and how long can it exist in the atmosphere or above. A binary companion sufficiently close to Mira A could reduce the local effective gravity enough to enhance mass loss and it could also account for the symmetric structures commonly observed in planetary nebulae.  The known companion Mira B is too far away to provide a gravity reduction by more than $\simeq$ 1\%.  A much closer companion would be needed, even a massive planet, but there is no evidence for another very close companion at this point. We find the dust clump to be on the side opposite to Mira B as  \citet{Chandler2007} did for the excess of dust emission measured with ISI. \citet{Khouri2018} did not find a systematic asymmetry at larger distance. They found a dust trail connecting Mira A and Mira B; however, there is no clear evidence for an influence of Mira B on the location of sites for dust production.

The size of the clump is larger than convective cells, such as those simulated in 3D hydrodynamical codes. However, the collective impact of complex convective structure could produce larger scale irregularities. The coupling between pulsations and convection could be invoked to explain such a clump of dust. A pioneer study of this coupling was performed by \citet{Anand&Michalitsanos1976} with a mechanical analog of a giant or a supergiant star. Their conclusion was that strong interactions can be expected with asymmetrical fluctuations of the surface brightness. \citet{Kiss2000} invoked  this mechanism as a possible explanation for  period and amplitude changes revealed in photometric monitoring. A single pulsation with effective convective coupling in the vicinity of a single or few large convective cells could well enhance the efficiency of mass ejection into the envelope, generating an over-density.  This over-density would presumably follow a ballistic trajectory, with much of the material falling back toward the photosphere, where it would encounter the next levitating shock.  It is easy to suppose that this pattern could preserve the over-density for some time, until it is reduced by leakage up, especially into the \edit{extended molecular layers}, with a timeframe that is consistently similar to the pulsational period.

\citet{Freytag&Hofner2008} performed 3D numerical simulations of an AGB star combining convection, pulsation, and dust formation which confirm the \citet{Anand&Michalitsanos1976} predictions. In their simulations, the bright convective cells provide a strongly asymmetric brightness distribution at and above levels they describe to be the top of the convective envelope which we believe corresponds approximately to what we have referred to as the core, or the depth to which near-IR imagery can penetrate.  Asymmetries in the density distribution appear most strongly in the highest levels, which are detected in the near-IR through their partial transparency and molecular absorption, which vary through the pulsational period. \citet{Freytag&Hofner2008} note a typical lifetime for convective structure of years, compared with a pulsational time of ~one year.  Something that could prove particularly significant for the upper envelope opacity, \citet{Freytag&Hofner2008} predicted localized dust formation in regions of enhanced density and depressed temperature which occur naturally in the complex flows and shocks. This has been confirmed by \citet{Hofner2019} with more recent 3D simulations of atmospheric shock waves generated by large-scale convective flows and pulsations. They find that the clumpy environment of dust can form around M-type AGB stars. This is clearly compatible with what we report here and with the result of \citet{Ohnaka2016,Adam2019,Norris2012}. This is also compatible with the dimming of the visual photometry at the time of our observations and during the few cycles that came before. 



\section{Conclusion and outlook}
\label{Sec:future}
The Mira image we obtained in H band at high spatial resolution offers supporting evidence for a possible outcome of pulsation-convection coupling and offers a possible direction for pursuing the evidence further. 

Confirmation should be possible with a series of interferometric images obtained with spectral information and polarimetry, possibly combined with visible resolved polarimetry, as already shown in some works cited in this paper. The three requirements include: excellent image quality, a combination between spectral and spatial resolution, and a suitable time series with a few month time interval over a few periods, that is, over several years. The quality of the images can be addressed in two ways. 

 First, for the higher spatial frequency baselines, good (u,v) coverage is essential. This is possible at present only at the VLTI, where the large selection of baselines could be optimized for a target list, although the need for rather contemporaneous UV coverage may be a challenge. Second, as is clearly shown by the aperture masking and other results mentioned above, these stars have a lot of low spatial frequency power.  It is essential to measure this correctly in order to properly interpret the high frequency measurements. This can be provided in the visible or near-infrared with extreme adaptive optics as shown by \citet{Ohnaka2016} with SPHERE or by aperture masking like the SAM mode of SPHERE on the VLT \citep{Cheetham2016} and with MICADO on the ELT in the future \citep{Lacour2014}, or with short-exposure imaging with adaptive optics in a moderate speckle mode \citep{Kervella2009}. Mid-infrared wavelengths can be reached without adaptative optics \citep{Poncelet2007} in this same mode. Near-IR polarimatric measurements are needed to simultaneously constrain scattering and absorption effects of dust. Spectral resolution is required to constrain temperatures and to separate the continuum (or pseudo-continuum) from the regions of the spectrum dominated by molecular opacity. Spectral resolutions as high as 4000 can be obtained with GRAVITY on the VLTI \citep{1stlight} in the K band. With such data, we could envisage the detection of a correlation between the brightness distribution on the photosphere and the evolving opacity in the envelope over subsequent years. The \citet{Freytag&Hofner2008} models specifically predict episodes of dust formation in the envelope, as close as 1-2 stellar radii from the surface. The combination of continuum and molecular opacities will help distinguish between dust and other continuous opacities. Some preliminary results on correlations between the visual light curve and the diameter measured in different band passes in the K band were recently published by \citet{Wittkowski2018} using GRAVITY. In addition, these data could be combined with mid-infrared four-telescope observations with the MATISSE instrument on VLTI \citep{Lopez2014} to further characterize the properties of the dust.

\begin{acknowledgements}
This research has made use of NASA's Astrophysics Data System Bibliographic Services, the SIMBAD and AFOEV databases, operated at CDS, Strasbourg, France and of the AAVSO database. The work of STR is supported by NOIRLab, which is managed by the Association of Universities for Research in Astronomy (AURA) under a cooperative agreement with the National Science Foundation. \edit{We thank the anonymous referee for suggestions which improved the content and presentation of the report.}



\end{acknowledgements}

\begin{appendix}
\section{Image reconstruction tests}
\label{Sec:AppA}

\begin{figure*}
\begin{tabular}{cc}
     \vspace{-0.24cm} 
     \includegraphics[width=\columnwidth]{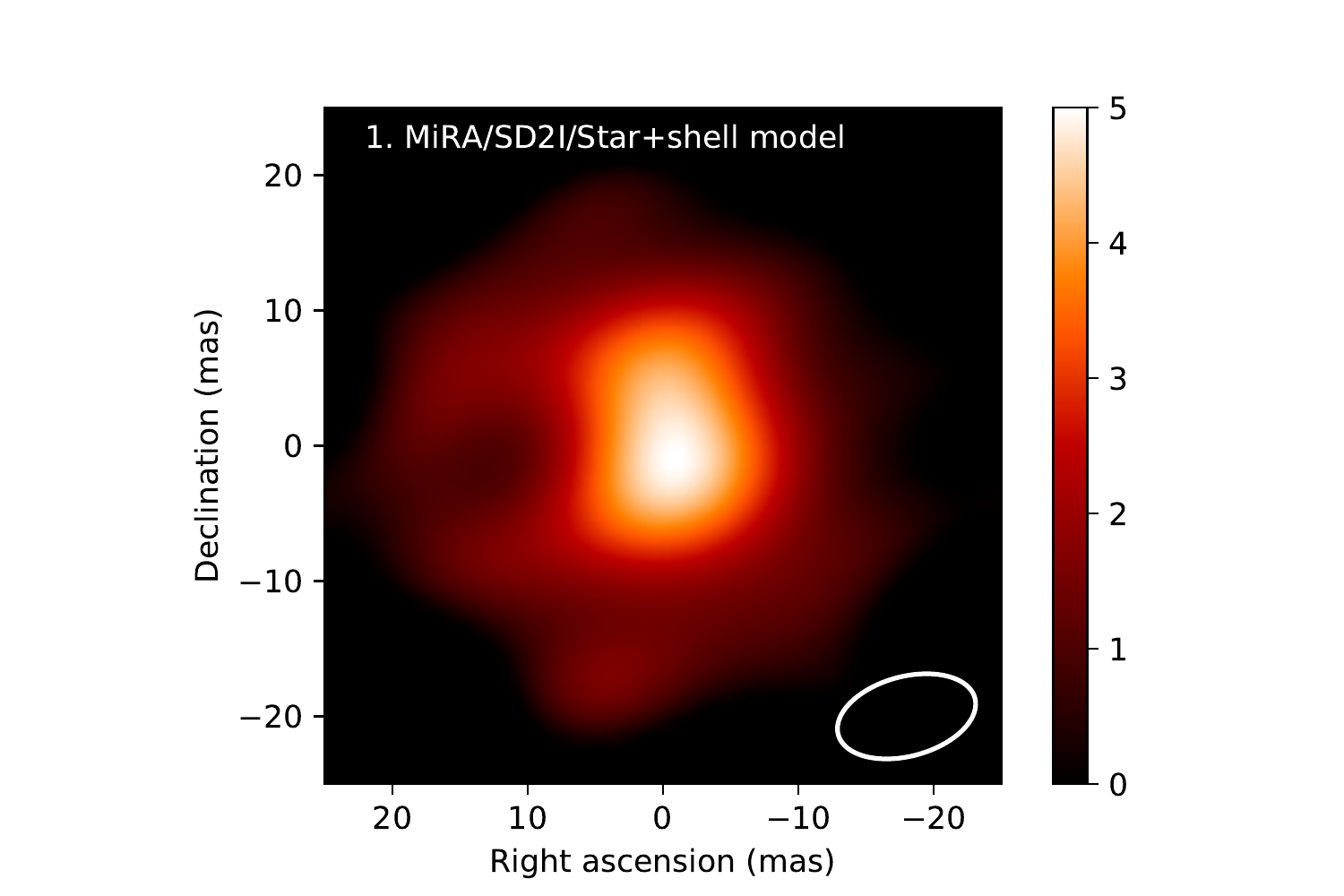}&
     \includegraphics[width=\columnwidth]{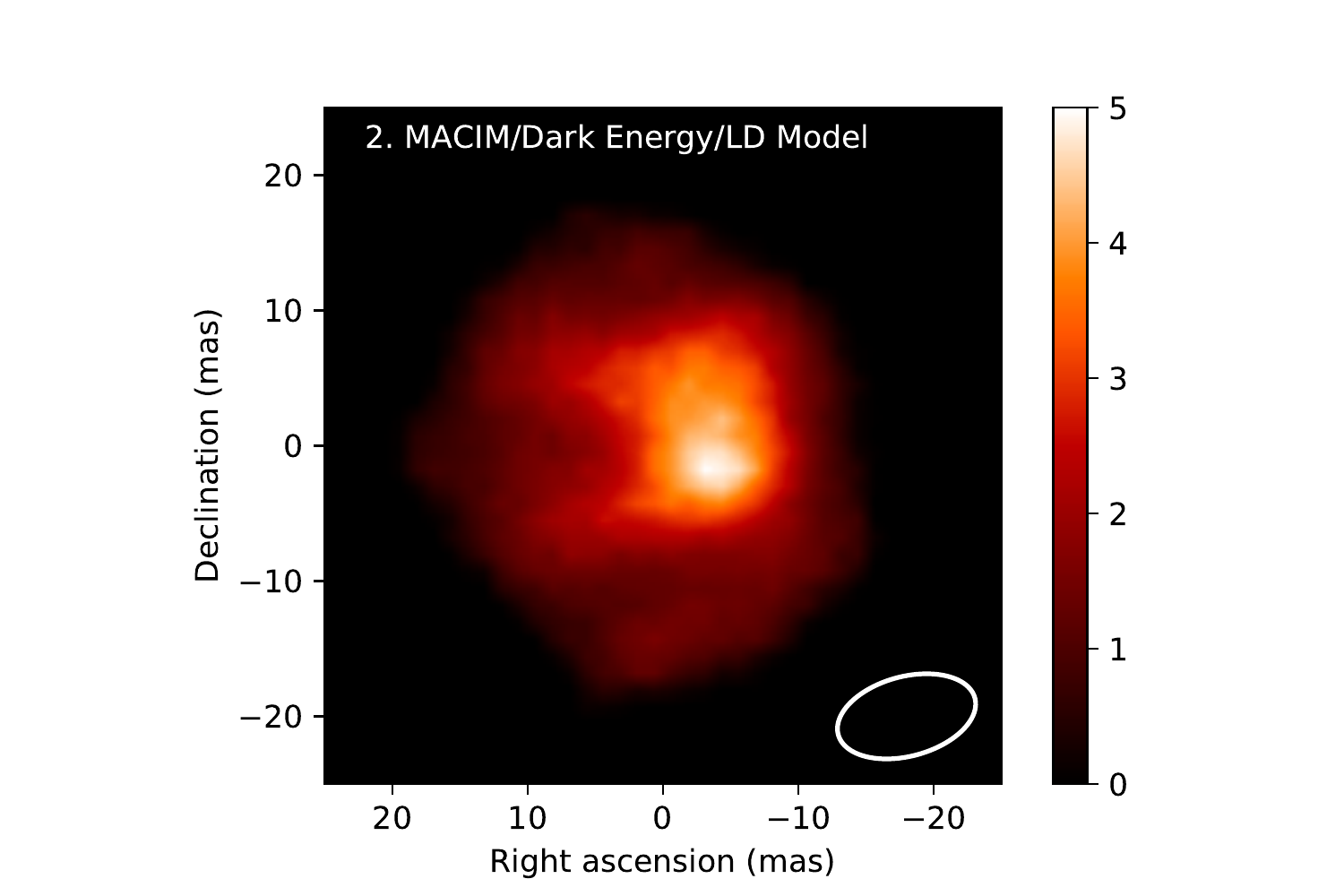}\\
       \vspace{-0.5cm}
    \includegraphics[width=\columnwidth]{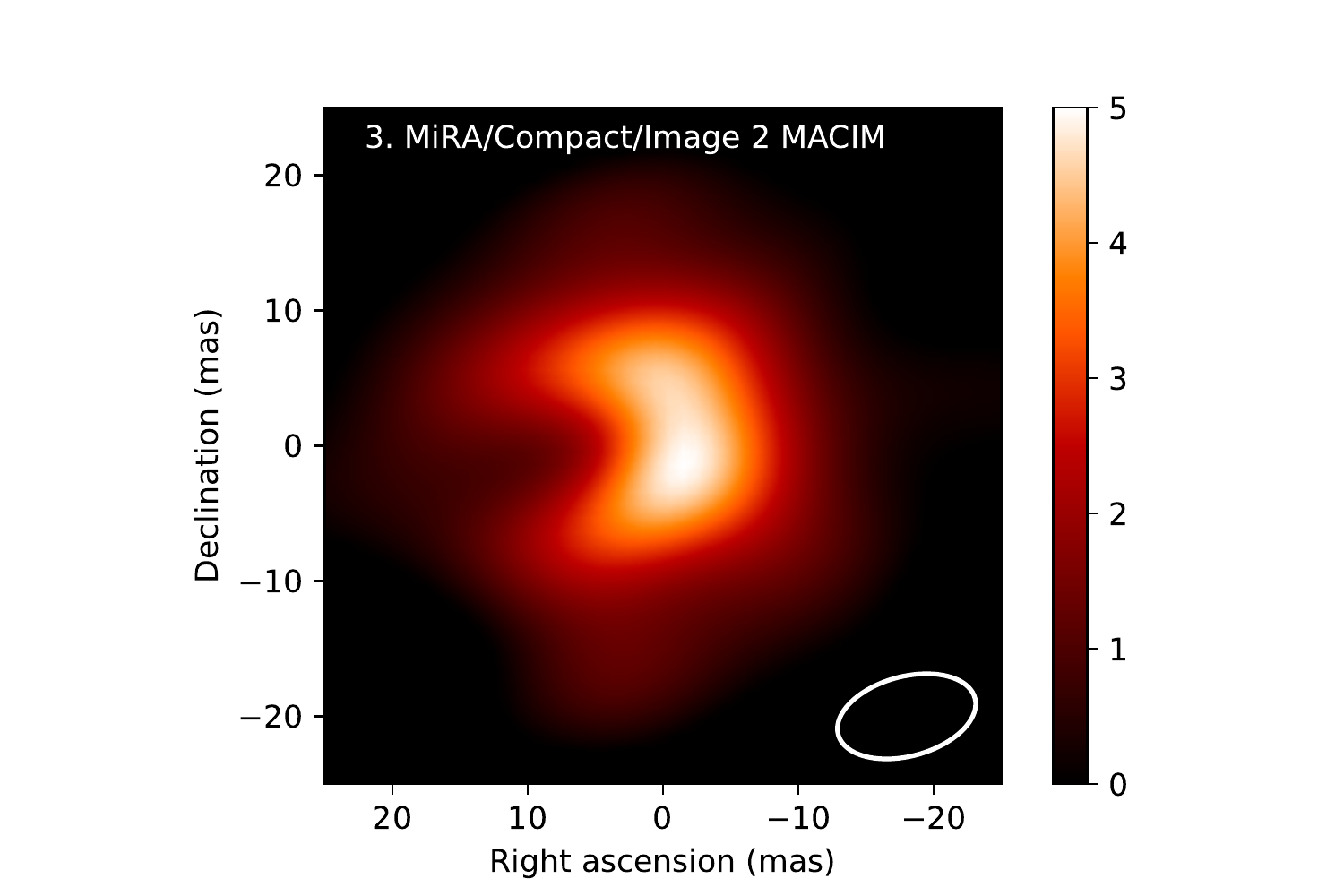} & 
    \includegraphics[width=\columnwidth]{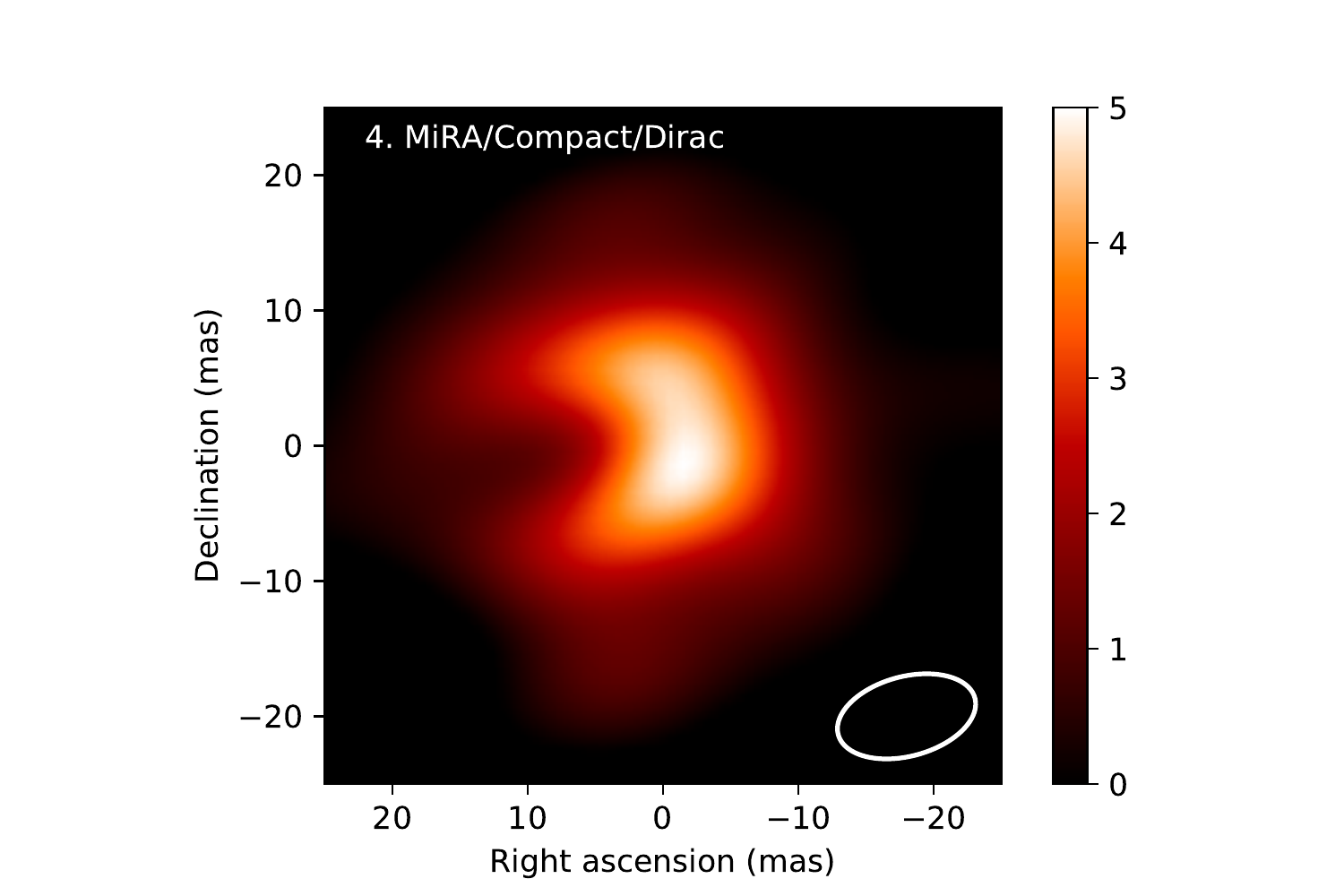} \\
    \includegraphics[width=\columnwidth]{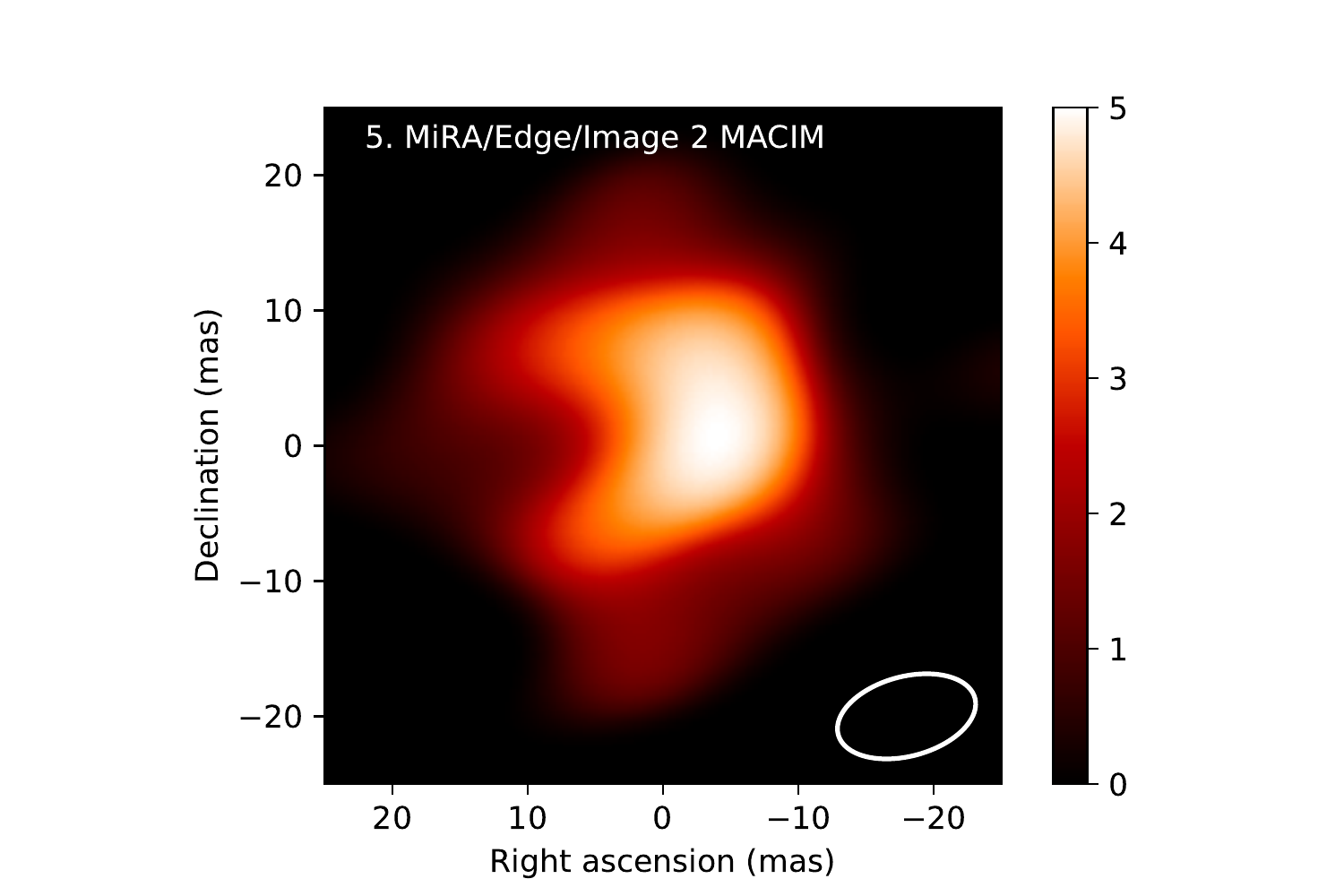} &  \includegraphics[width=\columnwidth]{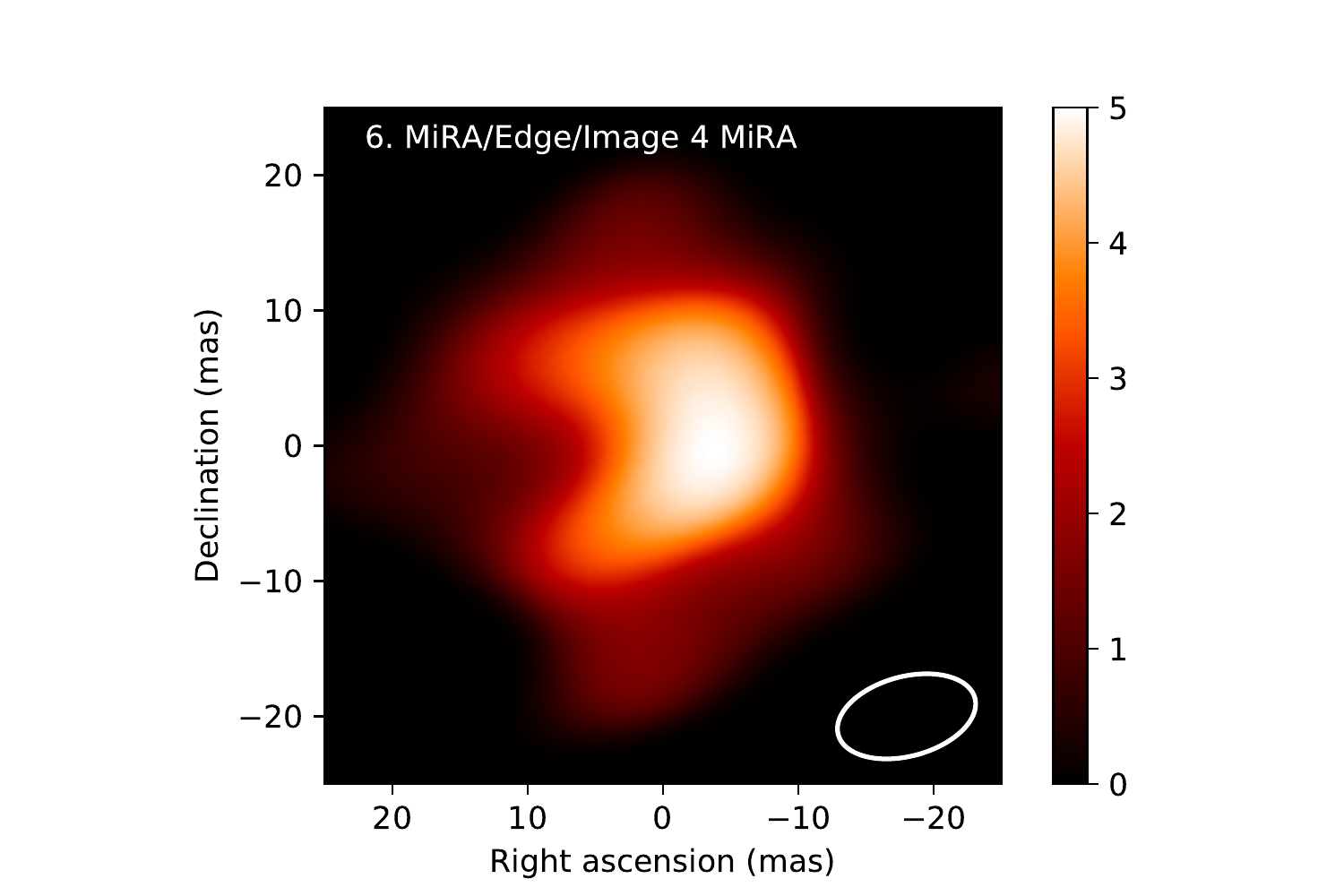} \\
\end{tabular}
\caption{\editguy{Images reconstructed with MACIM and MiRA. The name of the algorithm is given first followed by the type of regularization and the initial image. The ellipse at the bottom right is the FWHM of the central peak of the dirty beam.}}
\label{Fig:AppA}
\end{figure*}

\editguy{We tested two different softwares  with various priors and initial images to reconstruct the image of Mira: MACIM \citep{Ireland2006} and MiRA \citep{Thiebaut2008}.  }
\editguy{We define the reconstructed image $\V x \in \Omega \subset \Reals^n_+$, with
$\Omega$ the set of non-negative and normalized $n$-pixel images, as the one
that minimizes an objective function $\Cost(\V x)$.  In this case, the
objective function is the sum of three terms:}
\editguy{\begin{equation}
  \Cost(\V x) = \Cost_\Tag{V^2}(\V x) + \Cost_\Tag{cp}(\V x)
  + \mu\,\Cost_\Tag{rgl}(\V x) 
,\end{equation}}
\noindent\editguy{where $\Cost_\Tag{V^2}(\V x)$ and $\Cost_\Tag{cp}(\V x)$ are two data fidelity
terms measuring the distance between the model and the data (respectively for
the squared visibilities and for the closures phases) while $\Cost_\Tag{rgl}(\V
x)$ is a regularization term introduced to impose additional a priori\emph{}
constraints (so-called "priors" for short).  With MiRA, the hyper-parameter, $\mu > 0,$
is used to adjust the impact of the priors on the result.  For MACIM, the hyper-parameter is set to $\mu=1$. The data fidelity
terms are weighted according to the variance of the measurements. The $\chi^2_{\mathrm{data}}$ term in Sect.~4.1 is $\Cost_\Tag{V^2}(\V x) +
\Cost_\Tag{cp}(\V x)$. Dividing by the number of degrees of freedom yields the reduced $\chi^2$. Starting from a
given initial image $\V x_\Tag{init}$, MiRA \citep{Thiebaut_Giovannelli-2010-interferometry,Thiebaut_Young-2017-tutorial} iteratively reduces $\Cost(\V
x)$ to find a minimum. MACIM is a Monte-Carlo Markov chain algorithm who searches for the global minimum of $\Cost(\V
x)$. MACIM uses a simulated annealing algorithm with the Metropolis sampler. For the kind of considered data, the objective function
is not convex and the minimum of $\Cost(\V x)$ over $\Omega$ may not be unique. The result produced by MiRA and MACIM thus depend on the initial image, $\V x_\Tag{init}$, and on the regularization settings, $\mu$ and
$\Cost_\Tag{rgl}(\V x)$.}
\\

\begin{table*}[t]
\caption{Algorithms, initial images and regularizations used to reconstruct the 6 images of Fig.~\ref{Fig:AppA}}
    \centering
    \begin{tabular}{llllcc}
    \hline
    \hline
    \# & Algorithm & $\V x_\Tag{init}$ & $\Cost_\Tag{rgl}(\V x)$ & $\chi^{2}_{r,1}$ & $\chi^{2}_{r,2}$ \\
    \hline
    1 & MiRA & circularly symmetric star+layer model\tablefootmark{1} & $\Cost_\Tag{SD2I}(\V x)$, $\mu = 10^8$ & 8.214 & 1.088 \\
    2 & MACIM & circular limb-darkened disk\tablefootmark{2} & $ \Cost_\Tag{Dark~Energy}(\V x)$, $\mu = 1$ & 11.40 & 1.348 \\
    3 & MiRA & image 2 reconstructed by MACIM &
  $\Cost_\Tag{Compact}(\V x)$, $\mu = 2\times10^6$, $\gamma = 18$\,mas & 6.734 & 0.8473 \\ 
    4 & MiRA & point-like source & $\Cost_\Tag{Compact}(\V x)$, $\mu = 2\times10^6$, $\gamma = 18$\,mas & 6.734 & 0.8473 \\
    5 & MiRA & image 2 reconstructed by MACIM & $\Cost_\Tag{Edge}(\V x)$, $\mu = 10^4$, $\tau = 10^{-4}$ & 6.504 & 0.8362 \\
    6 & MiRA & image 4 reconstructed by MiRA & $\Cost_\Tag{Edge}(\V x)$, $\mu = 10^4$, $\tau = 10^{-4}$ & 6.503 & 0.8354 \\
    \hline
    \end{tabular}
    \tablefoot{
   \tablefoottext{1}{Circular star+layer model of Sect.~\ref{Sec:model} fitted to the $V^2$ data.}
   \tablefoottext{2}{Circular limb-darkened disk model fitted to the $V^2$ data.}
   }
    \label{tab:reconstructed_images}
\end{table*}

\noindent\editguy{For the reconstructions with MiRA, we considered three different regularizations.
The first one, which measures the squared distance of the image to the initial image is given by:}
\editguy{\begin{equation}
  \Cost_\Tag{SD2I}(\V x) = \sum_{j=1}^{n}
  {\Vert x_{j}-x_{init,j}\Vert^2}
.\end{equation}}
\editguy{The second one, favors compact images:}
\editguy{\begin{equation}
  \Cost_\Tag{Compact}(\V x) = \sum_{j=1}^{n}
  \bigl(1 + (2\,\Vert\V\theta_j\Vert/\gamma)^2\bigr)\,x_j^2
,\end{equation}}
\editguy{with $\V\theta_j$ the 2-dimensional angular position of pixel $j$ relative to
the center of the field of view, $\gamma > 0$ the full width at half maximum
(FWHM) of the assumed prior image ($\V\theta_j$ and $\gamma$ being in the same
units). Attempting to minimize $\Cost_\Tag{Compact}(\V x)$ for $\V x \in
\Omega$ imposes that the brightest parts of the sought image be mostly
concentrated around the center of the field of view.  Hence this regularization
corresponds to a "compactness"\ prior \citep{Thiebaut_Young-2017-tutorial}. The third one is given by:}
\editguy{\begin{equation}
  \Cost_\Tag{Edge}(\V x) = \sum_{j=1}^{n}
  \sqrt{\Vert\nabla_{\!j}\V x\Vert^2 + \tau^2}
,\end{equation}}
\editguy{where $\nabla_{\!j}\V x$ denotes the 2D spatial gradient of $\V x$
at pixel $j$ and $\tau > 0$ is a selected threshold.  This kind of
regularization penalizes the image gradients and thus imposes a smooth
image. The penalization is however less severe for the image gradients larger
than $\tau$ in magnitude than for the image gradients smaller than $\tau$ in
magnitude which helps preserving the edges in the image.  Hence this
regularization imposes an edge-preserving smoothness \emph{} prior.}

\editguy{For MACIM, the regularizer $\Cost_\Tag{Dark~Energy}(\V x)$ is a dark interaction energy regularizer. This regularizer is the sum of all pixel boundaries with zero flux on either side of the pixel boundary. It gives higher weight to large regions of dark space in between regions of flux.}

\editguy{In addition to the image presented in Fig.~\ref{Fig:image} and reconstructed with the SD2I regularization and an intial image obtained by fitting a circularly symmetric star+ shell model to the $V^2$ data, we reconstructed four other images with MiRA and one with MACIM. These images are presented in Fig.~\ref{Fig:AppA} together with the image of Fig.~\ref{Fig:image}. The details for each image are listed in Table~\ref{tab:reconstructed_images}. Images of 100x100 pixels were reconstructed with MiRA while 40x40 pixels were used for MACIM. The pixel size is $1$\,mas for MiRA and 1.44\,mas for MACIM. As explained in Sect.~\ref{Sec:imaging}, different $\mu$ values were tried for the reconstruction of image 1 with MiRA and the best result was obtained for $\mu=10^8$. For images 3 to 6, also reconstructed with MiRA, the value of $\mu$ was adjusted to minimize the  $\chi^2$. The parameter $\tau$ is in units of intensity per pixel and therefore depends on the pixel size and on the level of the flux normalization \citep{Renard_et_al-2011-regularization}. For the image reconstructed with MACIM, $\mu$ was set to 1. }

\editguy{The reduced $\chi^2$ are given Table~\ref{tab:reconstructed_images}. The number of degrees of freedom is the number of independent parameters subtracted to the number of data (940, i.e. 705 $V^2$ and 235 closure phases). The number of independent parameters is not as easy to define as in classical $\chi^2$ fitting. The maximum number of independent parameters is the number of pixels, e.g. 10,000 for the images reconstructed with MiRA. This number is larger than the number of data but not all pixels are independent because 1) they oversample the angular resolution elements in the image and 2) the regularization induces correlations between pixels. What matters here are the relative values to compare the images, the absolute values of the reduced $\chi^2$ are therefore only indicative. We have chosen to define the reduced $\chi^2$ as: $\chi^2_r = \chi^2 / \#{\mathrm{data}} = \chi^2 /940$. Our definition of $\chi^2_r$ is thus an underestimation of the reduced $\chi^2$. The errors provided by the pipeline have been modified for image 2 reconstructed with MACIM: a 5\% relative error was applied to the $V^2$ data with an additive default floor of 0.0001 and a minimum error of $0.5^{\circ}$ was used for the closure phase data. Raw error bars were used for the images reconstructed with MiRA (1 and 4 to 6). We have computed the reduced $\chi^2$ in two different ways: for $\chi^2_{r,1}$ we compared the reconstructed image with the $V^2$ and closure phase data using the raw error bars while for $\chi^2_{r,2}$ the modified error bars were used. Values are given in Table~\ref{tab:reconstructed_images}. The lower values of the $\chi^2_{r,2}$ estimator are a good indicator that the error bars produced by the pipeline are underestimated as noted in Sect.~\ref{Sec:Dust_patch_model}. The reduced $\chi^2_{r,2}$ are all close to 1 meaning all images are reconstructed with the same quality with respect to the data.}

\begin{figure}
\includegraphics[width=\columnwidth]{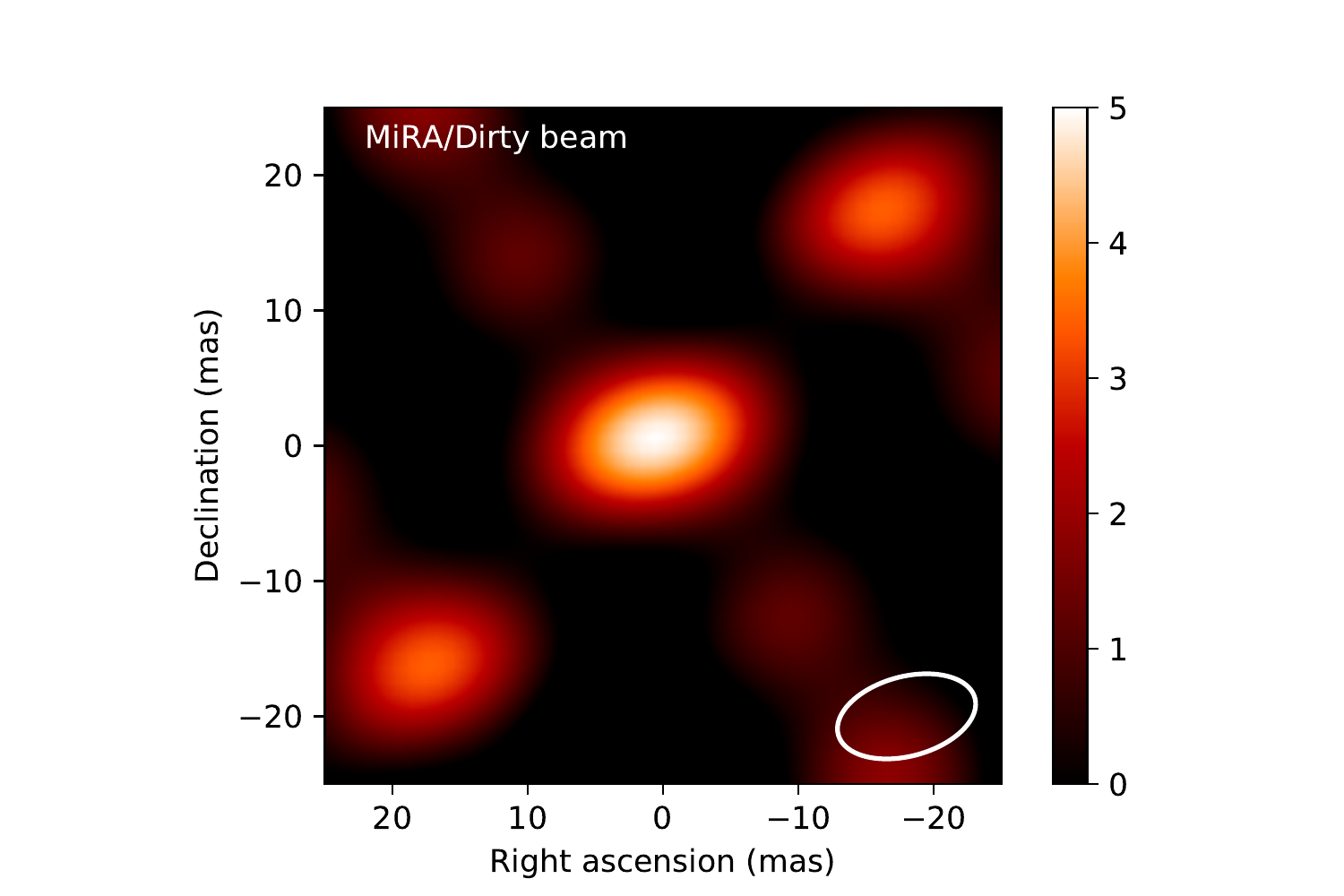}
\caption{\editguy{Dirty beam reconstructed with MiRA. The ellipse at the bottom right is the FWHM of the central peak. It gives the scale of the resolution in the reconstructed images.}}
\label{Fig:AppDB}
\end{figure}

\editguy{In addition we have reconstructed the dirty beam with MiRA (Fig.~\ref{Fig:AppDB}). All the visibilities of the sampled $(u,v)$ points were set to 1 for this image. The white ellipse on the image gives the FWHM of the central peak and gives the scale of the resolution in the reconstructed images.}

\editguy{In spite of the aforementioned issue of the solution not being unique, the
reconstructions in Fig.~\ref{Fig:AppA} clearly show that the result does not  strictly depend on the given algorithm that was used. We see no major influence of the initial image on the final image, in particular, there is no influence coming from a circular or absence of symmetry. The {\it Compact} regularization with MiRA makes the brightest peak appear a bit larger compared to the other reconstructions but does not change the image substantially. The features in the image cannot be explained by the structures of the dirty beam. Overall, the reconstructed images are very similar and exhibit the same remarkable structures with a $\simeq$20\,mas north-south elongated central object surrounded by a fainter $\simeq$40\,mas circular environment and a dark patch in the environment that is west of the bright central peak. We conclude that the image reconstruction process is quite robust despite the limited $(u,v)$ coverage. Image 1 featured in Sect.~\ref{Fig:image} is fairly representative of all the reconstructed images.}

\end{appendix}

%
%

\bibliographystyle{aa} 
\bibliography{Mira_GP_21_08_2020}

\begin{thebibliography}{65}
\expandafter\ifx\csname natexlab\endcsname\relax\def\natexlab#1{#1}\fi

\bibitem[{{Adam} \& {Ohnaka}(2019)}]{Adam2019}
{Adam}, C. \& {Ohnaka}, K. 2019, arXiv e-prints, arXiv:1907.05534

\bibitem[{{Anand} \& {Michalitsanos}(1976)}]{Anand&Michalitsanos1976}
{Anand}, S.~P.~S. \& {Michalitsanos}, A.~G. 1976, \apss, 45, 175

\bibitem[{{Aronson} {et~al.}(2017){Aronson}, {Bladh}, \&
  {H{\"o}fner}}]{Aronson2017}
{Aronson}, E., {Bladh}, S., \& {H{\"o}fner}, S. 2017, \aap, 603, A116

\bibitem[{{Baldwin} {et~al.}(1996){Baldwin}, {Beckett}, {Boysen}, {Burns},
  {Buscher}, {Cox}, {Haniff}, {Mackay}, {Nightingale}, {Rogers}, {Scheuer},
  {Scott}, {Tuthill}, {Warner}, {Wilson}, \& {Wilson}}]{Baldwin1996}
{Baldwin}, J.~E., {Beckett}, M.~G., {Boysen}, R.~C., {et~al.} 1996, \aap, 306,
  L13

\bibitem[{{Berger} {et~al.}(2003){Berger}, {Haguenauer}, {Kern},
  {Rousselet-Perraut}, {Malbet}, {Gluck}, {Lagny}, {Schanen-Duport}, {Laurent},
  {Delboulbe}, {Tatulli}, {Traub}, {Carleton}, {Millan-Gabet}, {Monnier},
  {Pedretti}, \& {Ragland}}]{Berger2003}
{Berger}, J.-P., {Haguenauer}, P., {Kern}, P.~Y., {et~al.} 2003, in \procspie,
  Vol. 4838, Interferometry for Optical Astronomy II, ed. W.~A. {Traub},
  1099--1106

\bibitem[{{Bladh} \& {H{\"o}fner}(2012)}]{Bladh2012}
{Bladh}, S. \& {H{\"o}fner}, S. 2012, \aap, 546, A76

\bibitem[{{Bladh} {et~al.}(2015){Bladh}, {H{\"o}fner}, {Aringer}, \&
  {Eriksson}}]{Bladh2015}
{Bladh}, S., {H{\"o}fner}, S., {Aringer}, B., \& {Eriksson}, K. 2015, \aap,
  575, A105

\bibitem[{{Bobrowsky}(2007)}]{Bobrowsky2007}
{Bobrowsky}, M. 2007, in Asymmetrical Planetary Nebulae IV, 22

\bibitem[{{Bohren} \& {Huffman}(1983)}]{Bohren&Huffman1983}
{Bohren}, C.~F. \& {Huffman}, D.~R. 1983, {Absorption and scattering of light
  by small particles} (New York: Wiley)

\bibitem[{{Borde} {et~al.}(2002){Borde}, {Coud{\'e} du Foresto}, {Chagnon}, \&
  {Perrin}}]{Borde2002}
{Borde}, P., {Coud{\'e} du Foresto}, V., {Chagnon}, G., \& {Perrin}, G. 2002,
  VizieR Online Data Catalog, 339

\bibitem[{{Burns} {et~al.}(1998){Burns}, {Baldwin}, {Boysen}, {Haniff},
  {Lawson}, {Mackay}, {Rogers}, {Scott}, {St. -Jacques}, {Warner}, {Wilson}, \&
  {Young}}]{Burns1998}
{Burns}, D., {Baldwin}, J.~E., {Boysen}, R.~C., {et~al.} 1998, \mnras, 297, 462

\bibitem[{{Chandler} {et~al.}(2007){Chandler}, {Tatebe}, {Wishnow}, {Hale}, \&
  {Townes}}]{Chandler2007}
{Chandler}, A.~A., {Tatebe}, K., {Wishnow}, E.~H., {Hale}, D.~D.~S., \&
  {Townes}, C.~H. 2007, \apj, 670, 1347

\bibitem[{{Cheetham} {et~al.}(2016){Cheetham}, {Girard}, {Lacour}, {Schworer},
  {Haubois}, \& {Beuzit}}]{Cheetham2016}
{Cheetham}, A.~C., {Girard}, J., {Lacour}, S., {et~al.} 2016, in Society of
  Photo-Optical Instrumentation Engineers (SPIE) Conference Series, Vol. 9907,
  \procspie, 99072T

\bibitem[{{Coud{\'e} du Foresto} {et~al.}(1997){Coud{\'e} du Foresto},
  {Ridgway}, \& {Mariotti}}]{Coude1997}
{Coud{\'e} du Foresto}, V., {Ridgway}, S., \& {Mariotti}, J.-M. 1997, \aaps,
  121, 379

\bibitem[{{Creech-Eakman} \& {Thompson}(2009)}]{Creech-Eakman&Thompson2009}
{Creech-Eakman}, M.~J. \& {Thompson}, R.~R. 2009, in Astronomical Society of
  the Pacific Conference Series, Vol. 412, The Biggest, Baddest, Coolest Stars,
  ed. D.~G. {Luttermoser}, B.~J. {Smith}, \& R.~E. {Stencel}, 149

\bibitem[{{Freytag} \& {H{\"o}fner}(2008)}]{Freytag&Hofner2008}
{Freytag}, B. \& {H{\"o}fner}, S. 2008, \aap, 483, 571

\bibitem[{{Freytag} {et~al.}(2017){Freytag}, {Liljegren}, \&
  {H{\"o}fner}}]{Freytag2017}
{Freytag}, B., {Liljegren}, S., \& {H{\"o}fner}, S. 2017, \aap, 600, A137

\bibitem[{{Gravity Collaboration} {et~al.}(2017){Gravity Collaboration},
  {Abuter}, {Accardo}, {Amorim}, {Anugu}, {{\'A}vila}, {Azouaoui}, {Benisty},
  {Berger}, {Blind}, {Bonnet}, {Bourget}, {Brandner}, {Brast}, {Buron},
  {Burtscher}, {Cassaing}, {Chapron}, {Choquet}, {Cl{\'e}net}, {Collin},
  {Coud{\'e} Du Foresto}, {de Wit}, {de Zeeuw}, {Deen},
  {Delplancke-Str{\"o}bele}, {Dembet}, {Derie}, {Dexter}, {Duvert}, {Ebert},
  {Eckart}, {Eisenhauer}, {Esselborn}, {F{\'e}dou}, {Finger}, {Garcia}, {Garcia
  Dabo}, {Garcia Lopez}, {Gendron}, {Genzel}, {Gillessen}, {Gonte}, {Gordo},
  {Grould}, {Gr{\"o}zinger}, {Guieu}, {Haguenauer}, {Hans}, {Haubois}, {Haug},
  {Haussmann}, {Henning}, {Hippler}, {Horrobin}, {Huber}, {Hubert}, {Hubin},
  {Hummel}, {Jakob}, {Janssen}, {Jochum}, {Jocou}, {Kaufer}, {Kellner},
  {Kendrew}, {Kern}, {Kervella}, {Kiekebusch}, {Klein}, {Kok}, {Kolb}, {Kulas},
  {Lacour}, {Lapeyr{\`e}re}, {Lazareff}, {Le Bouquin}, {L{\`e}na}, {Lenzen},
  {L{\'e}v{\^e}que}, {Lippa}, {Magnard}, {Mehrgan}, {Mellein}, {M{\'e}rand},
  {Moreno-Ventas}, {Moulin}, {M{\"u}ller}, {M{\"u}ller}, {Neumann}, {Oberti},
  {Ott}, {Pallanca}, {Panduro}, {Pasquini}, {Paumard}, {Percheron}, {Perraut},
  {Perrin}, {Pfl{\"u}ger}, {Pfuhl}, {Phan Duc}, {Plewa}, {Popovic}, {Rabien},
  {Ram{\'{\i}}rez}, {Ramos}, {Rau}, {Riquelme}, {Rohloff}, {Rousset},
  {Sanchez-Bermudez}, {Scheithauer}, {Sch{\"o}ller}, {Schuhler}, {Spyromilio},
  {Straubmeier}, {Sturm}, {Suarez}, {Tristram}, {Ventura}, {Vincent},
  {Waisberg}, {Wank}, {Weber}, {Wieprecht}, {Wiest}, {Wiezorrek}, {Wittkowski},
  {Woillez}, {Wolff}, {Yazici}, {Ziegler}, \& {Zins}}]{1stlight}
{Gravity Collaboration}, {Abuter}, R., {Accardo}, M., {et~al.} 2017, \aap, 602,
  A94

\bibitem[{{Hinkle} \& {Lebzelter}(2015)}]{hinkle2015}
{Hinkle}, K.~H. \& {Lebzelter}, T. 2015, in EAS Publications Series, Vol.~71,
  EAS Publications Series, 249--250

\bibitem[{{Hinkle} {et~al.}(1984){Hinkle}, {Scharlach}, \& {Hall}}]{Hinkle1984}
{Hinkle}, K.~H., {Scharlach}, W.~W.~G., \& {Hall}, D.~N.~B. 1984, \apjs, 56, 1

\bibitem[{{H{\"o}fner}(2008)}]{Hofner2008}
{H{\"o}fner}, S. 2008, \aap, 491, L1

\bibitem[{{H{\"o}fner} {et~al.}(2016){H{\"o}fner}, {Bladh}, {Aringer}, \&
  {Ahuja}}]{Hofner2016}
{H{\"o}fner}, S., {Bladh}, S., {Aringer}, B., \& {Ahuja}, R. 2016, \aap, 594,
  A108

\bibitem[{{H{\"o}fner} \& {Freytag}(2019)}]{Hofner2019}
{H{\"o}fner}, S. \& {Freytag}, B. 2019, \aap, 623, A158

\bibitem[{{Ireland} {et~al.}(2006){Ireland}, {Monnier}, \&
  {Thureau}}]{Ireland2006}
{Ireland}, M.~J., {Monnier}, J.~D., \& {Thureau}, N. 2006, in Society of
  Photo-Optical Instrumentation Engineers (SPIE) Conference Series, Vol. 6268,
  \procspie, 62681T

\bibitem[{{J{\"a}ger} {et~al.}(2003){J{\"a}ger}, {Il'in}, {Henning},
  {Mutschke}, {Fabian}, {Semenov}, \& {Voshchinnikov}}]{Jaeger2003}
{J{\"a}ger}, C., {Il'in}, V.~B., {Henning}, T., {et~al.} 2003, \jqsrt, 79, 765

\bibitem[{{Kami{\'n}ski} {et~al.}(2017){Kami{\'n}ski}, {M{\"u}ller}, {Schmidt},
  {Cherchneff}, {Wong}, {Br{\"u}nken}, {Menten}, {Winters}, {Gottlieb}, \&
  {Patel}}]{Kaminski2017}
{Kami{\'n}ski}, T., {M{\"u}ller}, H.~S.~P., {Schmidt}, M.~R., {et~al.} 2017,
  \aap, 599, A59

\bibitem[{{Kervella} {et~al.}(2009){Kervella}, {Verhoelst}, {Ridgway},
  {Perrin}, {Lacour}, {Cami}, \& {Haubois}}]{Kervella2009}
{Kervella}, P., {Verhoelst}, T., {Ridgway}, S.~T., {et~al.} 2009, \aap, 504,
  115

\bibitem[{{Khouri} {et~al.}(2016){Khouri}, {Maercker}, {Waters}, {Vlemmings},
  {Kervella}, {de Koter}, {Ginski}, {De Beck}, {Decin}, {Min}, {Dominik},
  {O'Gorman}, {Schmid}, {Lombaert}, \& {Lagadec}}]{Khouri2016}
{Khouri}, T., {Maercker}, M., {Waters}, L.~B.~F.~M., {et~al.} 2016, \aap, 591,
  A70

\bibitem[{{Khouri} {et~al.}(2018){Khouri}, {Vlemmings}, {Olofsson}, {Ginski},
  {De Beck}, {Maercker}, \& {Ramstedt}}]{Khouri2018}
{Khouri}, T., {Vlemmings}, W.~H.~T., {Olofsson}, H., {et~al.} 2018, \aap, 620,
  A75

\bibitem[{{Kiss} {et~al.}(2000){Kiss}, {Szatm{\'a}ry}, {Szab{\'o}}, \&
  {Mattei}}]{Kiss2000}
{Kiss}, L.~L., {Szatm{\'a}ry}, K., {Szab{\'o}}, G., \& {Mattei}, J.~A. 2000,
  \aaps, 145, 283

\bibitem[{{Knapp}(1985)}]{knapp1985}
{Knapp}, G.~R. 1985, \apj, 293, 273

\bibitem[{{Labeyrie} {et~al.}(1977){Labeyrie}, {Koechlin}, {Bonneau}, {Blazit},
  \& {Foy}}]{Labeyrie1977}
{Labeyrie}, A., {Koechlin}, L., {Bonneau}, D., {Blazit}, A., \& {Foy}, R. 1977,
  \apjl, 218, L75

\bibitem[{{Lacour} {et~al.}(2014){Lacour}, {Baudoz}, {Gendron}, {Boccaletti},
  {Galicher}, {Cl{\'e}net}, {Gratadour}, {Buey}, {Rousset}, {Hartl}, \&
  {Davies}}]{Lacour2014}
{Lacour}, S., {Baudoz}, P., {Gendron}, E., {et~al.} 2014, in Society of
  Photo-Optical Instrumentation Engineers (SPIE) Conference Series, Vol. 9147,
  \procspie, 91479F

\bibitem[{{Lacour} {et~al.}(2008){Lacour}, {Meimon}, {Thi{\'e}baut}, {Perrin},
  {Verhoelst}, {Pedretti}, {Schuller}, {Mugnier}, {Monnier}, {Berger},
  {Haubois}, {Poncelet}, {Le Besnerais}, {Eriksson}, {Millan-Gabet}, {Ragland},
  {Lacasse}, \& {Traub}}]{Lacour2008}
{Lacour}, S., {Meimon}, S., {Thi{\'e}baut}, E., {et~al.} 2008, \aap, 485, 561

\bibitem[{{Lacour} {et~al.}(2009){Lacour}, {Thi{\'e}baut}, {Perrin}, {Meimon},
  {Haubois}, {Pedretti}, {Ridgway}, {Monnier}, {Berger}, {Schuller},
  {Woodruff}, {Poncelet}, {Le Coroller}, {Millan-Gabet}, {Lacasse}, \&
  {Traub}}]{Lacour2009}
{Lacour}, S., {Thi{\'e}baut}, E., {Perrin}, G., {et~al.} 2009, \apj, 707, 632

\bibitem[{{Le Bouquin} {et~al.}(2009){Le Bouquin}, {Lacour}, {Renard},
  {Thi{\'e}baut}, {Merand}, \& {Verhoelst}}]{Le_Bouquin2009}
{Le Bouquin}, J.-B., {Lacour}, S., {Renard}, S., {et~al.} 2009, \aap, 496, L1

\bibitem[{{Liljegren} {et~al.}(2017){Liljegren}, {H{\"o}fner}, {Eriksson}, \&
  {Nowotny}}]{Liljegren2017}
{Liljegren}, S., {H{\"o}fner}, S., {Eriksson}, K., \& {Nowotny}, W. 2017, \aap,
  606, A6

\bibitem[{{Lopez} {et~al.}(1997){Lopez}, {Danchi}, {Bester}, {Hale}, {Lipman},
  {Monnier}, {Tuthill}, {Townes}, {Degiacomi}, {Geballe}, {Greenhill},
  {Cruzal{\`e}bes}, {Lef{\`e}vre}, {M{\'e}karnia}, {Mattei}, {Nishimoto}, \&
  {Kervin}}]{lopez1997}
{Lopez}, B., {Danchi}, W.~C., {Bester}, M., {et~al.} 1997, \apj, 488, 807

\bibitem[{{Lopez} {et~al.}(2014){Lopez}, {Lagarde}, {Jaffe}, {Petrov},
  {Sch{\"o}ller}, {Antonelli}, {Beckmann}, {Berio}, {Bettonvil}, {Glindemann},
  {Gonzalez}, {Graser}, {Hofmann}, {Millour}, {Robbe-Dubois}, {Venema}, {Wolf},
  {Henning}, {Lanz}, {Weigelt}, {Agocs}, {Bailet}, {Bresson}, {Bristow},
  {Dugu{\'e}}, {Heininger}, {Kroes}, {Laun}, {Lehmitz}, {Neumann}, {Augereau},
  {Avila}, {Behrend}, {van Belle}, {Berger}, {van Boekel}, {Bonhomme},
  {Bourget}, {Brast}, {Clausse}, {Connot}, {Conzelmann}, {Cruzal{\`e}bes},
  {Csepany}, {Danchi}, {Delbo}, {Delplancke}, {Dominik}, {van Duin}, {Elswijk},
  {Fantei}, {Finger}, {Gabasch}, {Gay}, {Girard}, {Girault}, {Gitton},
  {Glazenborg}, {Gont{\'e}}, {Guitton}, {Guniat}, {De Haan}, {Haguenauer},
  {Hanenburg}, {Hogerheijde}, {ter Horst}, {Hron}, {Hugues}, {Hummel},
  {Idserda}, {Ives}, {Jakob}, {Jasko}, {Jolley}, {Kiraly}, {K{\"o}hler},
  {Kragt}, {Kroener}, {Kuindersma}, {Labadie}, {Leinert}, {Le Poole}, {Lizon},
  {Lucuix}, {Marcotto}, {Martinache}, {Martinot-Lagarde}, {Mathar}, {Matter},
  {Mauclert}, {Mehrgan}, {Meilland}, {Meisenheimer}, {Meisner}, {Mellein},
  {Menardi}, {Menut}, {Merand}, {Morel}, {Mosoni}, {Navarro}, {Nussbaum},
  {Ottogalli}, {Palsa}, {Panduro}, {Pantin}, {Parra}, {Percheron}, {Duc},
  {Pott}, {Pozna}, {Przygodda}, {Rabbia}, {Richichi}, {Rigal}, {Roelfsema},
  {Rupprecht}, {Schertl}, {Schmidt}, {Schuhler}, {Schuil}, {Spang},
  {Stegmeier}, {Thiam}, {Tromp}, {Vakili}, {Vannier}, {Wagner}, \&
  {Woillez}}]{Lopez2014}
{Lopez}, B., {Lagarde}, S., {Jaffe}, W., {et~al.} 2014, The Messenger, 157, 5

\bibitem[{{Monnier}(2001)}]{Monnier2001}
{Monnier}, J.~D. 2001, \pasp, 113, 639

\bibitem[{{Monnier} {et~al.}(2006){Monnier}, {Berger}, {Millan-Gabet}, {Traub},
  {Schloerb}, {Pedretti}, {Benisty}, {Carleton}, {Haguenauer}, {Kern},
  {Labeye}, {Lacasse}, {Malbet}, {Perraut}, {Pearlman}, \&
  {Zhao}}]{Monnier2006}
{Monnier}, J.~D., {Berger}, J.-P., {Millan-Gabet}, R., {et~al.} 2006, \apj,
  647, 444

\bibitem[{{Norris} {et~al.}(2012){Norris}, {Tuthill}, {Ireland}, {Lacour},
  {Zijlstra}, {Lykou}, {Evans}, {Stewart}, \& {Bedding}}]{Norris2012}
{Norris}, B.~R.~M., {Tuthill}, P.~G., {Ireland}, M.~J., {et~al.} 2012, \nat,
  484, 220

\bibitem[{{Ohnaka} {et~al.}(2016){Ohnaka}, {Weigelt}, \&
  {Hofmann}}]{Ohnaka2016}
{Ohnaka}, K., {Weigelt}, G., \& {Hofmann}, K.-H. 2016, \aap, 589, A91

\bibitem[{{Ohnaka} {et~al.}(2017){Ohnaka}, {Weigelt}, \&
  {Hofmann}}]{Ohnaka2017}
{Ohnaka}, K., {Weigelt}, G., \& {Hofmann}, K.~H. 2017, \aap, 597, A20

\bibitem[{{Pedretti} {et~al.}(2005){Pedretti}, {Traub}, {Monnier},
  {Millan-Gabet}, {Carleton}, {Schloerb}, {Brewer}, {Berger}, {Lacasse}, \&
  {Ragland}}]{Pedretti2005}
{Pedretti}, E., {Traub}, W.~A., {Monnier}, J.~D., {et~al.} 2005, \ao, 44, 5173

\bibitem[{{Perrin}(2003)}]{Perrin2003}
{Perrin}, G. 2003, \aap, 398, 385

\bibitem[{{Perrin} {et~al.}(2015){Perrin}, {Cotton}, {Millan-Gabet}, \&
  {Mennesson}}]{Perrin2015}
{Perrin}, G., {Cotton}, W.~D., {Millan-Gabet}, R., \& {Mennesson}, B. 2015,
  \aap, 576, A70

\bibitem[{{Perrin} {et~al.}(2004){Perrin}, {Ridgway}, {Mennesson}, {Cotton},
  {Woillez}, {Verhoelst}, {Schuller}, {Coud{\'e} du Foresto}, {Traub},
  {Millan-Gabet}, \& {Lacasse}}]{Perrin2004}
{Perrin}, G., {Ridgway}, S.~T., {Mennesson}, B., {et~al.} 2004, \aap, 426, 279

\bibitem[{{Poncelet} {et~al.}(2007){Poncelet}, {Doucet}, {Perrin}, {Sol}, \&
  {Lagage}}]{Poncelet2007}
{Poncelet}, A., {Doucet}, C., {Perrin}, G., {Sol}, H., \& {Lagage}, P.~O. 2007,
  \aap, 472, 823

\bibitem[{{Ragland} {et~al.}(2006){Ragland}, {Traub}, {Berger}, {Danchi},
  {Monnier}, {Willson}, {Carleton}, {Lacasse}, {Millan-Gabet}, {Pedretti},
  {Schloerb}, {Cotton}, {Townes}, {Brewer}, {Haguenauer}, {Kern}, {Labeye},
  {Malbet}, {Malin}, {Pearlman}, {Perraut}, {Souccar}, \&
  {Wallace}}]{Ragland2006}
{Ragland}, S., {Traub}, W.~A., {Berger}, J.-P., {et~al.} 2006, \apj, 652, 650

\bibitem[{Renard {et~al.}(2011)Renard, Thiébaut, \&
  Malbet}]{Renard_et_al-2011-regularization}
Renard, S., Thiébaut, {\'E}., \& Malbet, F. 2011, \aap, 533, A64

\bibitem[{{Ryde} {et~al.}(2000){Ryde}, {Gustafsson}, {Eriksson}, \&
  {Hinkle}}]{Ryde2000}
{Ryde}, N., {Gustafsson}, B., {Eriksson}, K., \& {Hinkle}, K.~H. 2000, \apj,
  545, 945

\bibitem[{{Scholz}(2003)}]{Scholz2003}
{Scholz}, M. 2003, in \procspie, Vol. 4838, Interferometry for Optical
  Astronomy II, ed. W.~A. {Traub}, 163--171

\bibitem[{{Thi{\'e}baut}(2008)}]{Thiebaut2008}
{Thi{\'e}baut}, E. 2008, in \procspie, Vol. 7013, Optical and Infrared
  Interferometry, 70131I

\bibitem[{Thiébaut \&
  Giovannelli(2010)}]{Thiebaut_Giovannelli-2010-interferometry}
Thiébaut, {\'E}. \& Giovannelli, J.-F. 2010, IEEE Signal Processing Magazine,
  27, 97

\bibitem[{Thiébaut \& Young(2017)}]{Thiebaut_Young-2017-tutorial}
Thiébaut, {\'E}. \& Young, J. 2017, Journal of the Optical Society of America
  A, 34, 904

\bibitem[{{Traub}(1998)}]{Traub1998}
{Traub}, W.~A. 1998, in \procspie, Vol. 3350, Astronomical Interferometry, ed.
  R.~D. {Reasenberg}, 848--855

\bibitem[{{Tsuji}(2000)}]{Tsuji2000}
{Tsuji}, T. 2000, \apj, 538, 801

\bibitem[{{Whitelock} {et~al.}(2000){Whitelock}, {Marang}, \&
  {Feast}}]{Whitelock2000}
{Whitelock}, P., {Marang}, F., \& {Feast}, M. 2000, \mnras, 319, 728

\bibitem[{{Wittkowski} {et~al.}(2011){Wittkowski}, {Boboltz}, {Ireland},
  {Karovicova}, {Ohnaka}, {Scholz}, {van Wyk}, {Whitelock}, {Wood}, \&
  {Zijlstra}}]{Wittkowski2011}
{Wittkowski}, M., {Boboltz}, D.~A., {Ireland}, M., {et~al.} 2011, \aap, 532, L7

\bibitem[{{Wittkowski} {et~al.}(2016){Wittkowski}, {Chiavassa}, {Freytag},
  {Scholz}, {H{\"o}fner}, {Karovicova}, \& {Whitelock}}]{Wittkowski2016}
{Wittkowski}, M., {Chiavassa}, A., {Freytag}, B., {et~al.} 2016, \aap, 587, A12

\bibitem[{{Wittkowski} {et~al.}(2018){Wittkowski}, {Rau}, {Chiavassa},
  {H{\"o}fner}, {Scholz}, {Wood}, {de Wit}, {Eisenhauer}, {Haubois}, \&
  {Paumard}}]{Wittkowski2018}
{Wittkowski}, M., {Rau}, G., {Chiavassa}, A., {et~al.} 2018, \aap, 613, L7

\bibitem[{{Wong} {et~al.}(2016){Wong}, {Kami{\'n}ski}, {Menten}, \&
  {Wyrowski}}]{Wong2016}
{Wong}, K.~T., {Kami{\'n}ski}, T., {Menten}, K.~M., \& {Wyrowski}, F. 2016,
  \aap, 590, A127

\bibitem[{{Wood}(2007)}]{wood2007}
{Wood}, P.~R. 2007, in IAU Symposium, Vol. 239, Convection in Astrophysics, ed.
  F.~{Kupka}, I.~{Roxburgh}, \& K.~L. {Chan}, 343--348

\bibitem[{{Woodruff} {et~al.}(2008){Woodruff}, {Tuthill}, {Monnier}, {Ireland
  }, {Bedding}, {Lacour}, {Danchi}, \& {Scholz}}]{Woodruff2008}
{Woodruff}, H.~C., {Tuthill}, P.~G., {Monnier}, J.~D., {et~al.} 2008, \apj,
  673, 418

\end{thebibliography}
\end{document}